\documentclass[ twocolumn, times]{aastex631}

\usepackage{float}

\shorttitle{Hierarchical structures in Rosette}
\shortauthors{He et al.}
\graphicspath{{./}{figures/}}

\begin{document}

\title{Hierarchical Structure and Self-gravity in the Rosette Molecular Cloud}

\correspondingauthor{Yuehui Ma, Hongchi Wang}
\email{mayh@pmo.ac.cn, hcwang@pmo.ac.cn}

\author[0009-0004-2947-4020]{Suziye He}
\affiliation{Purple Mountain Observatory and Key Laboratory of Radio Astronomy, Chinese Academy of Sciences, 10 Yuanhua Road, Nanjing 210033, People's Republic of China\\}
\affiliation{School of Astronomy and Space Science, University of Science and Technology of China, 96 Jinzhai Road, Hefei 230026, People's Republic of China\\}

\author[0000-0002-8051-5228]{Yuehui Ma}
\affiliation{Purple Mountain Observatory and Key Laboratory of Radio Astronomy, Chinese Academy of Sciences, 10 Yuanhua Road, Nanjing 210033, People's Republic of China\\}

\author[0000-0003-0746-7968]{Hongchi Wang}
\affiliation{Purple Mountain Observatory and Key Laboratory of Radio Astronomy, Chinese Academy of Sciences, 10 Yuanhua Road, Nanjing 210033, People's Republic of China\\}
\affiliation{School of Astronomy and Space Science, University of Science and Technology of China, 96 Jinzhai Road, Hefei 230026, People's Republic of China\\}

\author[0000-0003-2254-1206]{Renjie Shen}
\affiliation{Purple Mountain Observatory and Key Laboratory of Radio Astronomy, Chinese Academy of Sciences, 10 Yuanhua Road, Nanjing 210033, People's Republic of China\\}

\author[0000-0002-6388-649X]{Miaomiao Zhang}
\affiliation{Purple Mountain Observatory and Key Laboratory of Radio Astronomy, Chinese Academy of Sciences, 10 Yuanhua Road, Nanjing 210033, People's Republic of China\\}

\author[0000-0003-2218-3437]{Chong Li}
\affiliation{Purple Mountain Observatory and Key Laboratory of Radio Astronomy, Chinese Academy of Sciences, 10 Yuanhua Road, Nanjing 210033, People's Republic of China\\}

\author[0009-0009-3431-1150]{Zhenyi Yue}
\affiliation{Purple Mountain Observatory and Key Laboratory of Radio Astronomy, Chinese Academy of Sciences, 10 Yuanhua Road, Nanjing 210033, People's Republic of China\\}
\affiliation{School of Astronomy and Space Science, University of Science and Technology of China, 96 Jinzhai Road, Hefei 230026, People's Republic of China\\}

\author[0009-0009-8707-8620]{Xiangyu Ou}
\affiliation{Purple Mountain Observatory and Key Laboratory of Radio Astronomy, Chinese Academy of Sciences, 10 Yuanhua Road, Nanjing 210033, People's Republic of China\\}
\affiliation{School of Astronomy and Space Science, University of Science and Technology of China, 96 Jinzhai Road, Hefei 230026, People's Republic of China\\}

\author[0000-0003-3151-8964]{Xuepeng Chen}
\affiliation{Purple Mountain Observatory and Key Laboratory of Radio Astronomy, Chinese Academy of Sciences, 10 Yuanhua Road, Nanjing 210033, People's Republic of China\\}
\affiliation{School of Astronomy and Space Science, University of Science and Technology of China, 96 Jinzhai Road, Hefei 230026, People's Republic of China\\}



\begin{abstract}

We analyze the hierarchical structure in the Rosette Molecular Cloud (RMC) using $\rm ^{13}CO\ J=1-0$ data from the Milky Way Imaging Scroll Painting (MWISP) survey with a non-binary Dendrogram algorithm that allows multiple branches to emerge from parent structures. A total of 588 substructures are identified, including 458 leaves and 130 branches. The physical parameters of the substructures, including peak brightness temperature ($T_{\text{peak}}$), brightness temperature difference ($T_{\text{diff}}$), radius ($R$), mass ($M$), velocity dispersion ($\sigma_v$), and surface density ($\Sigma$), are characterized. The $T_{\text{peak}}$ and $T_{\text{diff}}$ distributions follow exponential functions with characteristic values above $5\ \sigma_{\rm RMS}$. The statistical properties and scaling relations, i.e., $\sigma_v-R$, $M-R$, and $\sigma_v-R\Sigma$ relations are in general consistent with those from traditional segmentation methods. The mass and radius follow power-law distributions with exponents of 2.2–2.5, with slightly flatter slopes for substructures inside the \ion{H}{2} region. The velocity dispersion scales weakly with radius ($\sigma_v \propto R^{0.45\pm0.03}$, $r=0.58$), but shows a tighter correlation with the product of surface density and size ($\sigma_v \propto (\Sigma R)^{0.29\pm0.01}$, $r=0.73$). Self-gravitating substructures are found across scales from $\sim$0.2 to 10 pc, and nearly all structures with peak brightness above 4 K are gravitationally bound ($\alpha_{\rm vir}<2$). The fraction of bound structures increases with mass, size, and surface density, supporting the scenario of global hierarchical collapse (GHC) for the evolution of molecular clouds, in which molecular clouds and their substructures are undergoing multiscale collapse.

\end{abstract}



\section{Introduction} \label{sec1}


Molecular clouds are the coldest and densest phase of the interstellar medium (ISM), in which star formation takes place. Influenced by factors such as self-gravity, magnetic fields, and turbulence, molecular clouds exhibit complex and hierarchical internal structures. A central question in the field of star formation is to know which factor plays the predominant role in the evolution of molecular clouds. Molecular clouds were once considered to be dominated by self-gravity \citep{Goldreich1974}, therefore being in a state of global gravitational collapse. However, this  assumption is challenged by the observed low star formation rates (SFRs) of molecular clouds and the absence in observations of characteristic molecular line profiles that would indicate global gravitational collapse \citep{Zuckerman1974a, Zuckerman1974b}. In this situation, supporting mechanisms are needed for molecular clouds to prevent global collapse. As presented in \cite{Crutcher2012}, magnetic fields often fail to provide sufficient support on relatively large ($\sim$ pc) scales in typical molecular cloud environments, given that the mass-to-flux ratios of molecular clouds are generally supercritical. The relative importance between self-gravity and turbulence is still under debate \citep{Ballesteros2011a, Ballesteros2011b}. Some theories propose that turbulence governs the dynamics and structures of molecular clouds on a broad range of spatial scales, from tens of parsecs down to $\sim$pc, and that self-gravity starts to play a dominant role in the evolution of clumps and cores on sub-pc scales \citep{Elmegreen2004, McKee2007}. Nevertheless, the global hierarchical collapse (GHC) model proposed by \citet{Vazquez-Semadeni2019} once again emphasizes the substantial influence of self-gravity across all scales. In this framework, molecular clouds undergo gravitational collapse across various spatial scales: larger internal structures initiate contraction first, while later on smaller structures undergo faster contraction. Additionally, the observed supersonic linewidths of emission lines at millimeter wavelengths, often interpreted as observational evidence of turbulence, are considered the results of nonlinear asymmetrical collapse in the GHC model. Assessing the importance of self-gravity  with respect to turbulence at different spatial scales of molecular clouds are crucial for our understanding of star formation. 

Various methods can be employed to investigate the properties of turbulence and self-gravity in molecular clouds \citep{Burkhart2021}, including the probability distribution function (N-PDF) \citep{Kainulainen2011, Ma2020, Schneider2022}, structure function \citep{Chira2019, Henshaw2020}, principal component analysis (PCA) \citep{Heyer1997, Heyer2004, Heyer2006}, delta-variance \citep{Schneider2011, Elia2014}, Dendrogram \citep{Rosolowsky2008, Burkhart2013} and others. Among these methods, the N-PDF method focuses on the distribution of column density, thereby losing all spatial information. The structure function has diverse applications depending on the studied field, including investigating the effects of compressibility on turbulence scalings \citep{Kritsuk2007} and detecting anisotropy in observational data \citep{Esquivel2011}, among others. The delta-variance method quantifies structural features within a map at specific spatial scales by calculating the variance of its wavelet decomposition. PCA is a powerful dimensionality reduction technique that identifies correlated patterns based on the decomposition of the data covariance matrix; it is empirically applied to spectral line data cubes to extract velocity information through a pixel-by-pixel approach. Dendrogram is a widely used structure-identifying algorithm. It can decompose a three-dimensional (3D) position-position-velocity (PPV) data cube into interconnected structures and present the underlying connections between the structures with a tree-like diagram. It were used to focus exclusively on certain structural levels of the interstellar medium \citep{Rice2016, Cheng2018}, or to examine ISM properties across different spatial scales \citep{Goodman2009, Storm2014, Oakes2025}. Specifically, \cite{Goodman2009} applied the Dendrogram algorithm to analyze the $^{13}$CO emission line data of the L1448 star-forming region and found that gravity plays a crucial role over the full range of possible scales in their data, which is up to $\sim$1 pc. However, the role of self-gravity from $\sim$ pc to the giant molecular cloud (GMC) scale ($\rm \sim$ 10 pc) still need to be explored. 

On spatial scales from molecular clouds down to cores, statistical relations between different physical properties have been found. For example, \citet{Larson1981} identified three scaling relations: (1) the power-law relation, $\sigma_v \propto L^{0.38}$, between velocity dispersion, $\sigma_v$, and the spatial scale, $L$, of molecular clouds; (2) $2GM/\sigma_v^2 L  \sim 1$, indicating that molecular clouds are in self-gravitational equilibrium; and (3) the relationship between the number density and size of molecular clouds, i.e., $n \sim L^{-1.1}$, implying a constant surface density for molecular clouds. The power-law index of 0.38 in the $\sigma_v-L$ relationship is comparable to the expected value for incompressible turbulence, which is 1/3. Later studies, however, revealed that the linewidth–size relation is not universal. Many works report an exponent of about 0.5 \citep{Solomon1987, Heyer2004, Rice2016, Miville2017}, consistent with supersonic Burgers-type turbulence (e.g., \citealt{McKee2007}), while studies of massive star-forming regions often find shallower slopes or no correlation \citep{Caselli1995, Shirley2003, Ballesteros2011a, Traficante2018b}. Moreover, when variations in column density are considered, the simple $\sigma_v$–$L$ relation breaks down, and a surface-density–dependent form, $\sigma_v^2/R \propto \Sigma$, emerges instead \citep{Heyer2009}. 
In practice, most of the studies are based on cloud samples located in different environments or cloud structures obtained using traditional structural segmentation algorithms such as CLUMPFIND \citep{Williams1994}, GAUSSCLUMPS \citep{Stutzki1990}, and FELLWALKER \citep{Berry2015}.
Recently, we investigated the self-gravity and scaling relations for the hierarchical substructures of the Maddalena giant molecular cloud \citep{Shen2024}. The Maddalena cloud is a typical quiescent cloud, devoid of OB stars and with very little low-mass star formation, while the Rosette molecular cloud (RMC), on the other hand, represents active and high-mass star forming regions. In this work, we extend our analysis to the RMC.

The Rosette molecular cloud, located at a distance of 1.46 kpc \citep{Yan2019} in the third quadrant of the Galactic plane, is a well-known GMC with a mass of $\sim10^5\ M_\odot$, exhibiting abundant ongoing star formation \citep{Williams1995}. It is associated with the Rosette Nebula \citep{Roman2008b}, which is an \ion{H}{2} region driven by the NGC 2244 OB cluster. At the interface between the Rosette nebula and the RMC, the molecular cloud gas is illuminated and potentially compressed by the nebula \citep{Schneider1998}. The interface region exhibits complex velocity structure, relatively high excitation temperature, and substantial $\rm H_2$ column density. In comparison, regions outside the boundary of the \ion{H}{2} region, show relatively lower excitation temperature and $\rm H_2$ column density. It is still controversial whether star formation in the RMC is triggered by the expanding H II region \citep{Roman2008a} or not \citep{Schneider2012, Cambresy2013}, making it important to investigate the internal structure of the cloud. The RMC was previously observed in the $\rm ^{12}CO$ and $\rm ^{13}CO\ J=1-0$ lines by the Five College Radio Astronomy Observatory (FCRAO) 14 m telescope with an angular resolution of 47$\arcsec$ and a spectral resolution of $\rm \sim 0.13 \ km\ s^{-1}$ \citep{Heyer2006}, and further studied by \citet{Veltchev2018}. More recently, \citet{Li2018} presented complete $\rm ^{12}CO$, $\rm ^{13}CO$, and $\rm C^{18}O\ J=1-0$ observations of the RMC using the 13.7 m millimeter-wavelength telescope of the Purple Mountain Observatory (PMO-13.7 m) as part of the MWISP project \citep{Su2019}. Compared with the FCRAO data, the MWISP survey provides higher sensitivity, wider spatial coverage, and additional $\rm C^{18}O$ measurements, and and thus constitutes the dataset analyzed in this work.

In this study, we use the $^{13}$CO data from the MWISP survey to investigate the physical properties of the internal substructures in the RMC identified with the Dendrogram algorithm. Then, we discuss the role of gravity across scales in the RMC. Section \ref{sec2} introduces the data and algorithm used in this work. In Section \ref{sec3}, we present the statistics and scaling relations of the physical parameters, such as mass, radius, velocity dispersion, and others, for the substructures of the RMC. Finally, we discuss the importance of gravity at various spatial scales in RMC in Section \ref{sec4} and summarize our results in Section \ref{sec5}.

\section{Data and Method}\label{sec2} 
\subsection{Observation}\label{sec2.1}
\begin{figure*}[!htb]
	\centering 
	\includegraphics[width=0.8\textwidth]{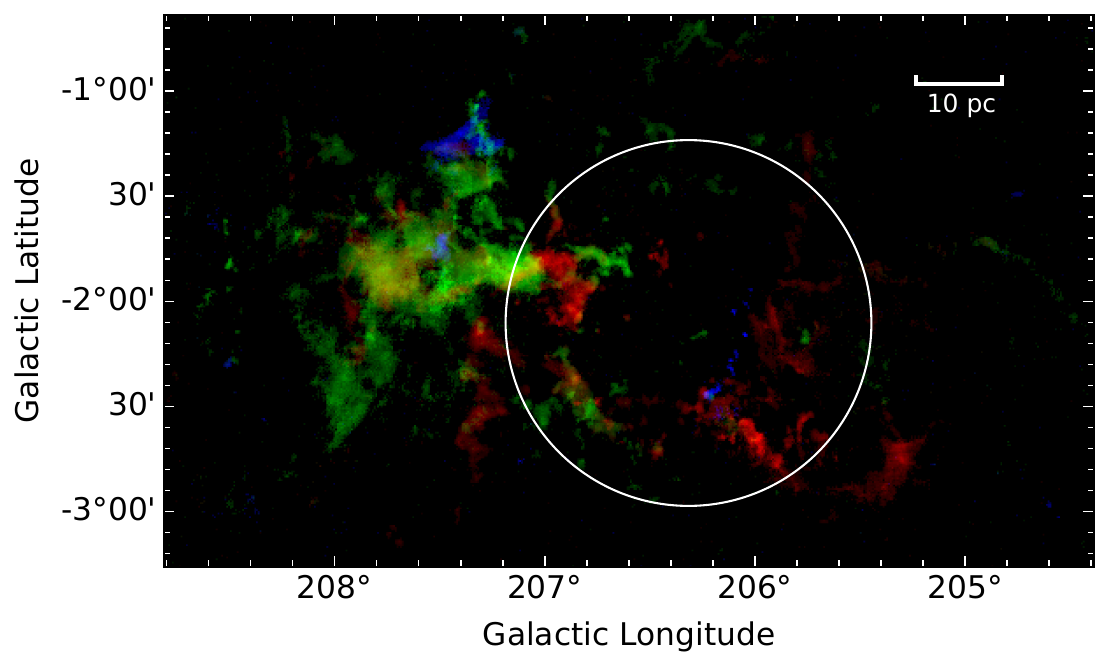}
	\caption{RGB map of the $^{13}$CO emission of the RMC. Blue, green, and red colors correspond to the integrated intensity of the $^{13}$CO emission in the velocity range of [$-$2, 5.5], [5.5, 13], and [13, 20.5] $\rm km\ s^{-1}$, respectively. The boundary of the Rosette Nebula given by \cite{Quireza2006} is overlaid with a white circle.}\label{fig1}
\end{figure*}

The RMC has been observed with the PMO-13.7 m telescope as part of the MWISP project, which is an unbiased large-scale survey of CO and its isotopologues toward the northern Galactic plane\citep{Su2019}. \cite{Li2018} presented the results toward a 3$.\!\!^{\circ}$5$\times$2$.\!\!^{\circ}$5 region, which was completed by the MWISP survey at that time. By the time of this study, the coverage of the RMC region had been expanded. We therefore adopt data from a larger area of 4$.\!\!^{\circ}$2$\times$2$.\!\!^{\circ}$7. Since \cite{Li2018} provided a detailed description of the observation, we only briefly introduce some key information in this work. The PMO-13.7 m telescope is equipped with a nine-beam sideband-separated superconducting spectroscopic array receiver (SSAR) system, enabling simultaneous observation of $\rm^{12}CO$, $\rm^{13}CO$, and $\rm C^{18}O\ J=1-0$ lines. The beam size of the $\rm ^{12}CO\ J=1-0$ line is $\sim 50^{\prime \prime}$ and our data are resampled to pixels of spatial size of $30^{\prime \prime} \times 30^{\prime \prime}$. The backend of the SSAR is a Fast Fourier Transform Spectrometer (FFTS) containing 16384 channels with a bandwidth of 1 GHz, which provides a velocity resolution of 0.16 km s$^{-1}$ at the $\rm ^{12}CO\ J=1-0$ frequency of 115 GHz. The final RMS noise levels, $\sigma_{\rm RMS}$, are approximately 0.5 K per 0.16 km s$^{-1}$ for $^{12}$CO and 0.3 K per 0.17 km s$^{-1}$ for $^{13}$CO.

We take $-$2 to 20.5 km s$^{-1}$ as the velocity span of the RMC to avoid unrelated components along the line of sight \citep{Li2018}. Compared to the $\rm^{12}CO$ emission, which is generally optically thick for molecular clouds,$\rm ^{13}CO\ J=1-0$ emission is relatively optically thin, making it a better tracer for internal substructures of molecular clouds. Therefore, our analyses mainly use the $\rm ^{13}CO\ J=1-0$ emission line data. Figure \ref{fig1} shows the color-coded $^{13}$CO integrated intensity map of the RMC with red, green, and blue colors representing integrated intensities within velocity ranges of [$-$2, 5.5], [5.5, 13], and [13, 20.5] km s$^{-1}$, respectively. The white circle in Figure \ref{fig1} indicates the location of the Rosette Nebula from the Wide-field Infrared Survey Explorer (WISE) \ion{H}{2} region catalog \citep{Anderson2014}. Further details of the nebula are presented in \cite{Quireza2006}. Complex and nested structures are obvious from Figure \ref{fig1}.
\subsection{The Non-binary Dendrogram}\label{sec2.2}

The astrodendro python package developed by \citet{Rosolowsky2008} is a widely used structure-identifying algorithm. 
It can decompose a 3D Position-Position-Velocity (PPV) data cube into interconnected structures, and present the underlying connections between the structures in a tree-like diagram. The algorithm starts by identifying local maxima in the dataset, and then progressively lowers the brightness temperature threshold, $T_{mb}$, defining 3D isosurfaces that enclose these maxima. When two isosurfaces intersect at a specific brightness temperature $T_{mb, merge}$, the original algorithm checks if the two structures enclosed by the two isosurfaces satisfy the independence criteria which are set by two input parameters, i.e., $min\_npix$ and $min\_delta$, of the algorithm. If the number of voxels contained in an isosurface is greater than $min\_npix$ and $T_{mb, maximum}-$ $T_{mb, merge}>min\_delta$, the structure is considered to be independent and labeled as a ``leaf''. The ``leaves'' merge to form ``branches'', and these ``branches'' can further merge with either ``leaves'' or other ``branches'' to construct a hierarchical ``tree''. This process continues until the $T_{mb}$ of the current isosurface reaches the input parameter, $min\_value$. The peak brightness temperature, $T_{\text{peak}}$, of the voxels contained within each structure is defined to be its ``height'', while the brightness temperature difference, $T_{\text{diff}}$, between the top and bottom ends of each structure is defined as its ``length''.

\begin{figure*}[!htb]
	\centering
	\begin{minipage}[t]{0.7\linewidth}
		\centering
		\includegraphics[width=\linewidth, clip]{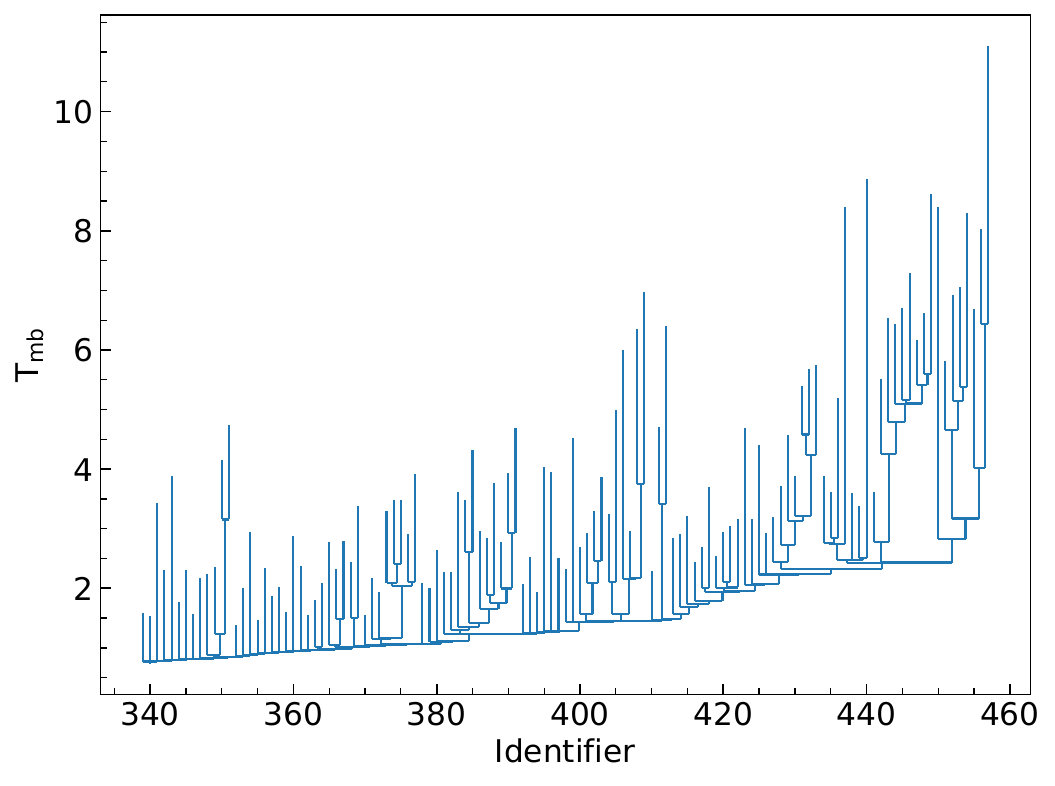}
		\\(a)
	\end{minipage}
	\begin{minipage}[t]{0.7\textwidth}
		\centering
		\includegraphics[width=\linewidth, clip]{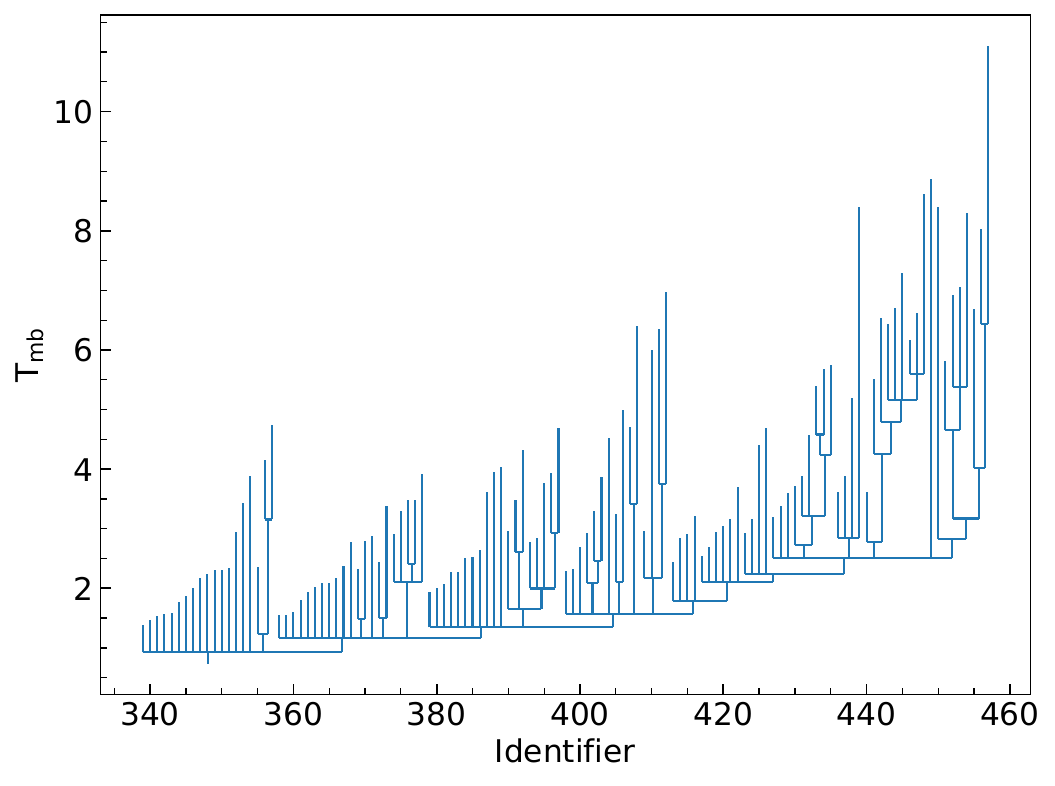}
		\\(b)
	\end{minipage}
	\caption{Examples of tree diagrams identified by (a) the original binary Dendrogram algorithm and (b) our modified non-binary Dendrogram algorithm.}
	\label{fig2}
\end{figure*}

In the original Dendrogram algorithm, each branch has only two direct children, resulting in a binary tree. Forcing binary mergers can produce artificial branches with $T_{\text{diff}}$ falling below the observational sensitivity, as illustrated by \cite{Storm2014} shown in their Figure 27. Therefore, we modified the original Dendrogram algorithm to eliminate branching caused by noise, i.e., branches with $T_{\text{diff}}$ below a threshold parameter,  $branch\_delta$, are considered dependent and are merged into a larger independent branch. Our modification allows more than two substructures to sprout from one branch. A comparison between the tree diagrams produced by the original binary Dendrogram algorithm and our modified non-binary version is shown in Figure \ref{fig2}. As shown in Figure \ref{fig2}, artificial branching in panel (a) is effectively removed in our modified non-binary Dendrogram algorithm. A similar non-binary Dendrogram algorithm was introduced by \citet{Storm2014}, albeit with a different technical implementation.

In the following analysis, the non-binary Dendrogram algorithm is applied to the $\rm ^{13}CO$ data with the parameter set of $min\_value=3\ \sigma_{\rm RMS}$, $min\_delta=2\ \sigma_{\rm RMS}$, $min\_npix=27$ voxels, and $branch\_delta=\sigma_{\rm RMS}$, where $\sigma_{\rm RMS}$ is the $^{13}$CO observational sensitivity. The requirement that one structure should contain at least 27 voxels corresponds to a minimum structure of size of $\rm \sim0.9\ pc\times0.9\ pc\times0.51\ km\ s^{-1}$, at the distance of the RMC. Branches and leaves that sprout directly from the $min\_ value$ threshold are called ``trunks''. The number of segments in the shortest path for a substructure linking to its ancestral trunk is defined as the ``level'' of that structure. For example, trunks have levels of 0, while their children have levels of 1, and so on. Details on the comparison of results from the original Dendrogram and the non-binary Dendrogram algorithms are given in Section \ref{appA1} in the Appendix. 

\subsection{Calculation of Physical Parameters}\label{sec2.3}
The column density of a structure is calculated assuming the CO molecules are under the local thermodynamic equilibrium (LTE) condition \citep{Dickman1978, Mangum2015}. The specific formula is as below, 

{\small
\begin{equation}
N(^{13}CO)=2.42\times10^{14}\frac{1+0.88/T_{ex}}{1-exp(-5.29/T_{ex})} T_{ex}\int\tau(\rm ^{13}CO)d\upsilon,\label{eq1}
\end{equation}
}where $T_{ex}$ is the excitation temperature derived from $\rm^{12}CO(1-0)$ emission and $\tau$ is the optical depth of $\rm^{13}CO(1-0)$ emission. When the optical depth of the $\rm^{13}CO(1-0)$ emission is small, the integral term $T_{ex}\int\tau(\rm ^{13}CO)d\upsilon$ in formula (\ref{eq1}) can be approximated to $\tau_0/(1-e^{-\tau_0})\int{T_{mb}dv}$ \citep{Pineda2010}, where $\tau_0$ is the peak optical depth of the $\rm^{13}CO(1-0)$ emission. The excitation temperature can be derived from the optically thick $\rm^{12}CO$ emission through

\begin{equation}
T_{ex}=\frac{5.53}{ln(1+\frac{5.53}{T_{peak}+0.819})},
\label{eq2}
\end{equation}
where $T_{peak}$ refers to the peak brightness temperature of $\rm^{12}CO(1-0)$ emission along each line of sight. The column density of $^{13}$CO can be converted to the $\rm H_2$ column density using the abundance ratio $[\rm H_2/^{12}CO] = 1.1 \times 10^4$ \citep{Frerking1982}, along with the isotopic ratio $\rm [^{12}\ C/^{13}C] = 6.21d_{\rm GC} + 18.71$ \citep{Milam2005}, where $d_{\rm GC}$ is the Galactocentric distance of the RMC in units of $\rm kpc$. The value of $d_{\rm GC}$ can be calculated from the heliocentric distance of the RMC, its Galactic longitude, and the Galactocentric radius of the Sun, which is assumed to be $R_\odot = 8.15$ kpc \citep{Reid2019}. Given the Galactocentric distance of the RMC (9.5 kpc), we adopt an isotopic ratio of $\rm [^{12}C/^{13}C] = 78$ for our analysis. By integrating the $\rm H_2$ column density over the solid angle occupied by a structure, we can obtain its total mass as

\begin{equation}
M=\mu m_HD^2\int N(\rm H_2)d\Omega,
\label{eq3}
\end{equation}
where $\mu = 2.8$ is the mean molecule weight, $m_H$ is the mass of a hydrogen atom, and $D=1.46\ \rm kpc$ is the distance to the RMC.

The angular radius of a structure, as defined by the Dendrogram algorithm, is proportional to the geometric mean of the intensity-weighted second moments along its major and minor axes. It can be expressed as follows:
\begin{equation}
R_{a}=\eta \sqrt{\sigma_{maj}\sigma_{min}}\label{eq4},
\label{eq4}
\end{equation}
where ${\sigma_{maj}}$ and ${\sigma_{min}}$ stand for rms size along major and minor axes, respectively.
We use the parameter $\eta=1.91$, as recommended by \citeauthor{Rosolowsky2008}, to correct the geometric mean radius of an ellipsoid to the effective radius of a sphere. The angular radius is converted to the physical radius using the distance to the RMC through $R=R_a\times D$. 

The virial parameter is defined as the absolute value of the ratio of twice the kinetic energy to the gravitational potential energy of a molecular cloud structure \citep{Bertoldi1992},

\begin{equation}
\alpha_{\text{vir}}=\frac{5\sigma_v^2R}{GM}, 
\label{eq5}
\end{equation}
where $G$ is the gravitational constant. The critical value $\alpha_{\text{vir}} = 1$ indicates that molecular clouds are under simple virial equilibrium, $2|E_k| = |E_g|$. In the case of pressure-bounded isothermal spheres, such as Bonnor-Ebert spheres \citep{Bonnor1956}, the mass threshold for maintaining hydrostatic equilibrium is 
\begin{equation}
    M_{BE} = 2.43 \frac{\sigma^2_v R}{G}.
\label{eq6}
\end{equation}
Combining formulae \ref{eq5} and \ref{eq6}, one can tell that the critical value of $\alpha_{\text{vir}}$ for BE spheres is $\sim$2. Clumps with $\alpha_{\text{vir}} < 2$ are considered to be gravitationally bound and may undergo gravitational collapse, whereas those with $\alpha_{\text{vir}} > 2$ either require external pressure to maintain equilibrium \citep{Bertoldi1992} or are dispersing.

\section{Results}\label{sec3}
\begin{figure*}[!htb]
	\centering
	\begin{minipage}[t]{0.8\linewidth}
		\centering
		\includegraphics[width=\linewidth, clip]{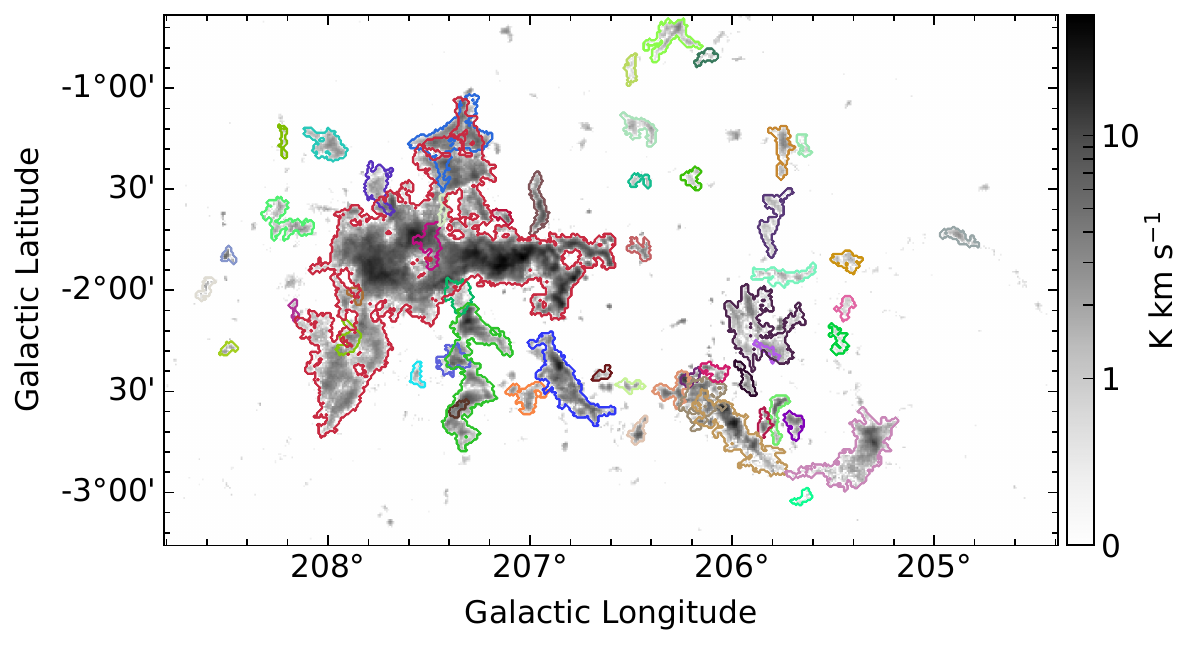}
		\\(a)
	\end{minipage}
	\begin{minipage}[t]{0.7\textwidth}
		\centering
		\includegraphics[width=\linewidth, clip]{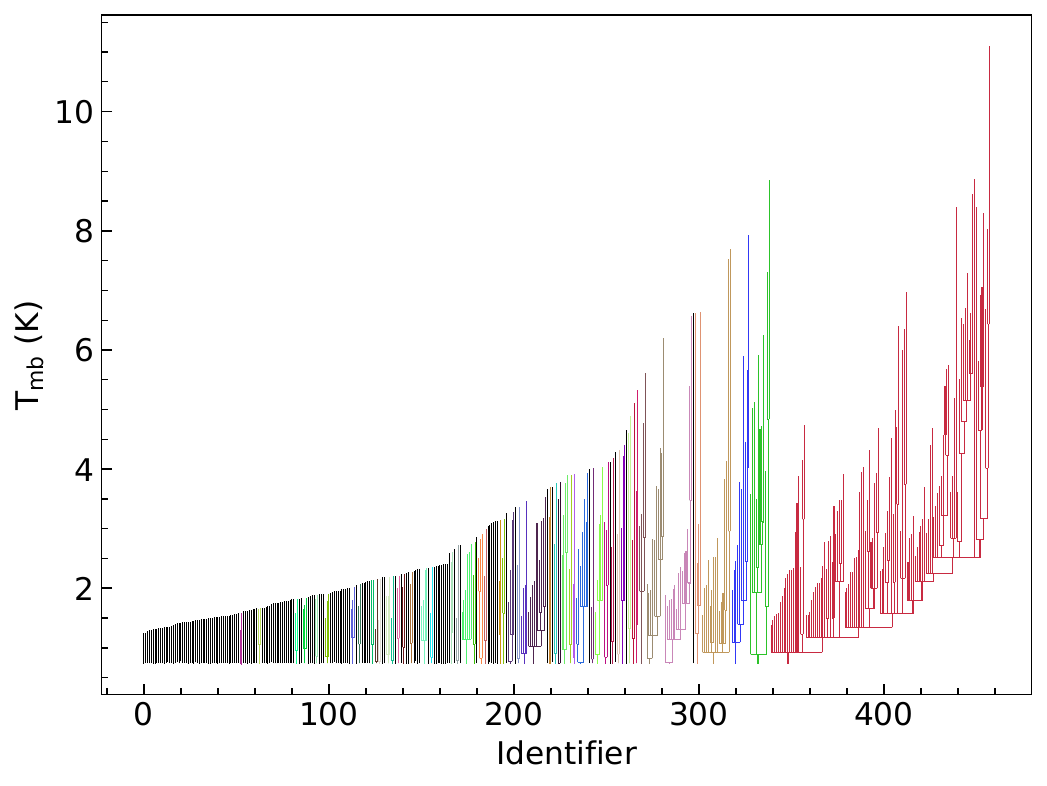}
		\\(b)
	\end{minipage}
	\caption{(a) Integrated intensity map of $^{13}$CO emission of the RMC within $-$2 to 20.5 km s$^{-1}$ overlaid with the projected boundaries of hierarchical trunks. The boundaries are colored the same way in panel (b).   (b) Tree diagram of the $\rm ^{13}CO$ emission of the RMC generated using the modified non-binary Dendrogram algorithm. The $x$ axis indicates the identifiers of each structure, which have no physical meaning, while the $y$ axis represents the brightness temperature spans of the structures. Hierarchical trunks are highlighted with different colors, while monadic trunks are marked in black.}\label{fig3}
\end{figure*}

Figure \ref{fig3} presents the tree diagram generated by the non-binary Dendrogram algorithm, which comprises 588 structures, including 458 leaves and 130 branches. In total, the identified structures constitute 196 trunks, and collectively they recover 86.7$\%$ of the total flux above the $3\ \sigma_{\rm RMS}$ threshold. We outline the boundaries of the hierarchical trunks, i.e., those with at least two layers of structures, projected onto the $l-b$ plane integrated intensity map in Figure \ref{fig3}(a), using different colors. The corresponding tree diagrams of the trunks are displayed in the same colors in Figure \ref{fig3}(b). In the remaining part of this section, we present statistics on the physical parameters of the substructures in the tree diagram. 

\subsection{Dendrogram Statistics}\label{sec3.1}

\begin{figure*}[!htb]
	\begin{minipage}[htbo]{\columnwidth}
		\centering
		\includegraphics[width=\columnwidth]{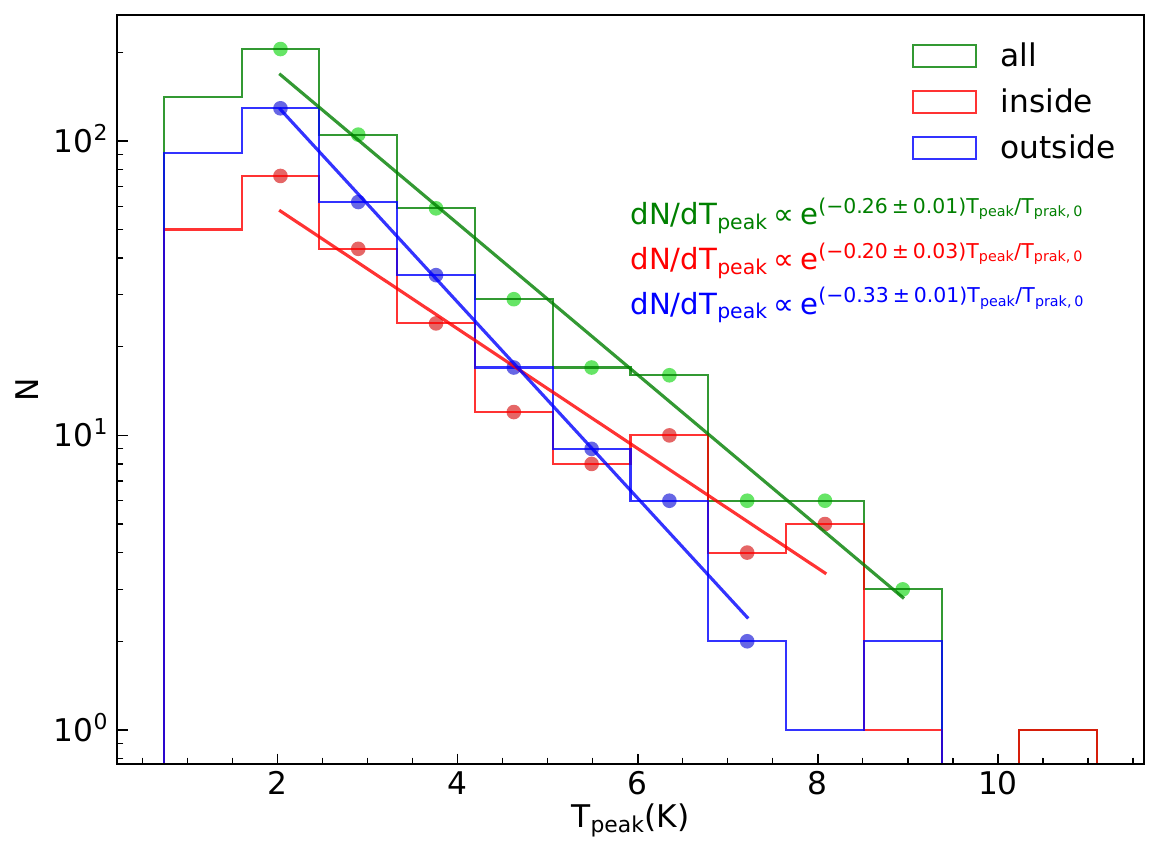}
		\\{(a)}
	\end{minipage}	
	\begin{minipage}[htbp]{\columnwidth}
		\centering
		\includegraphics[width=\columnwidth]{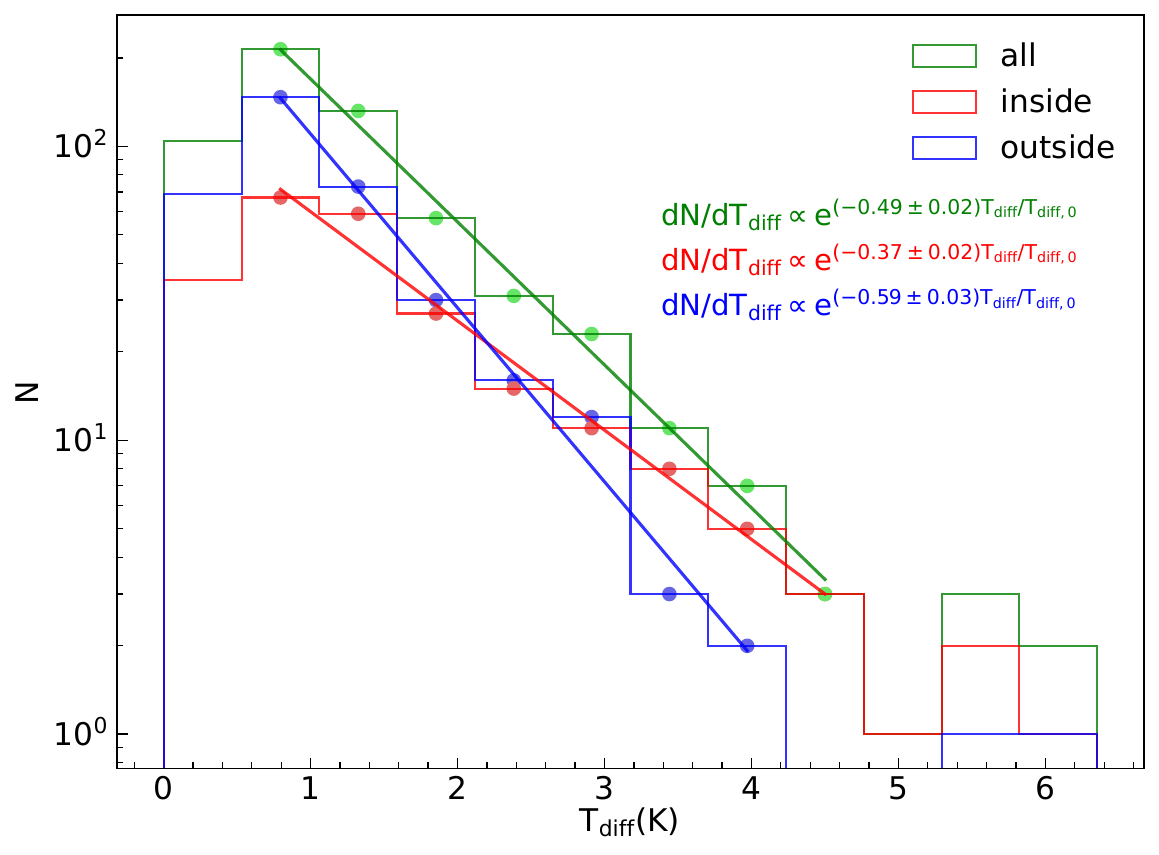}
		\\{(b)}
	\end{minipage}
\caption{ Histograms of (a) $T_{\text{peak}}$ ($h$) and (b) $T_{\text{diff}}$ ($l$) of the identified substructures. The green, red, and blue histograms correspond to all structures in the RMC, structures inside the Rosette nebula, and those outside the nebula, respectively. }\label{fig4}
\end{figure*}

The structures in the tree diagrams are mainly characterized by two physical properties, peak brightness temperature, $T_{\text{peak}}$, and brightness temperature difference, $T_{\text{diff}}$. Figure \ref{fig4} shows the histograms of the $T_{\text{peak}}$ and $T_{\text{diff}}$ of the structures.
The structures inside and outside the \ion{H}{2} region are presented in red color and blue color, respectively. The measured $T_{\text{peak}}$ and $T_{\text{diff}}$ exhibit exponential distributions for structures both inside and outside the \ion{H}{2} region. We fitted each histogram with an exponential function from its peak to the bin at the high end that contains at least two counts. Our fittings yield the relationships $\Delta N/\Delta T_{\text{peak}}\propto e^{(-0.20\pm0.02)T_{\text{peak}}/T_{\text{peak,0}}}$ for structures inside the \ion{H}{2} region and $\Delta N/\Delta T_{\text{peak}}\propto e^{(-0.32\pm0.01)T_{\text{peak}}/T_{\text{peak,0}}}$ for those outside the \ion{H}{2} region, where $T_{\text{peak,0}}$ denotes the characteristic $T_{\text{peak}}$ of the distribution. For structures inside the \ion{H}{2} region, $T_{\text{peak,0}}$ equals 5 K. For structures outside the \ion{H}{2} region, this value drops to 3.1 K. In terms of $T_{\text{diff}}$, the distributions yield $\Delta N/\Delta T_{\text{diff}}\propto e^{(-0.37\pm0.03)T_{\text{diff}}/T_{\text{diff,0}}}$ for structures inside the \ion{H}{2} region with $T_{\text{diff,0}}$ equal 2.7 K and $\Delta N/\Delta T_{\text{diff}}\propto e^{(-0.61\pm0.05)T_{\text{diff}}/T_{\text{diff,0}}}$ for those outside the \ion{H}{2} region with $T_{\text{diff,0}}$ equal to 1.6 K.

The typical $T_{\text{diff}}$ of the structures for both regions inside the outside the \ion{H}{2} region are significantly greater than the input parameter $min\_delta$, indicating that the structures of RMC traced by $\rm ^{13}$CO indeed exhibit intrinsic hierarchy. The slopes of the distributions inside the \ion{H}{2} region are shallower, suggesting a higher fraction of longer or higher structures in this region. This phenomenon could be attributed to feedback from the \ion{H}{2} region, which can heat and compress the surrounding molecular gas. This is consistent with the excitation temperature distribution in figure 16 of \citet{Li2018}, which shows that the molecular gas inside the \ion{H}{2} region is significantly warmer than outside. 

\subsection{Statistics of Physical Parameters} \label{sec3.2}
\begin{figure*}[!htb]
	\gridline{\fig{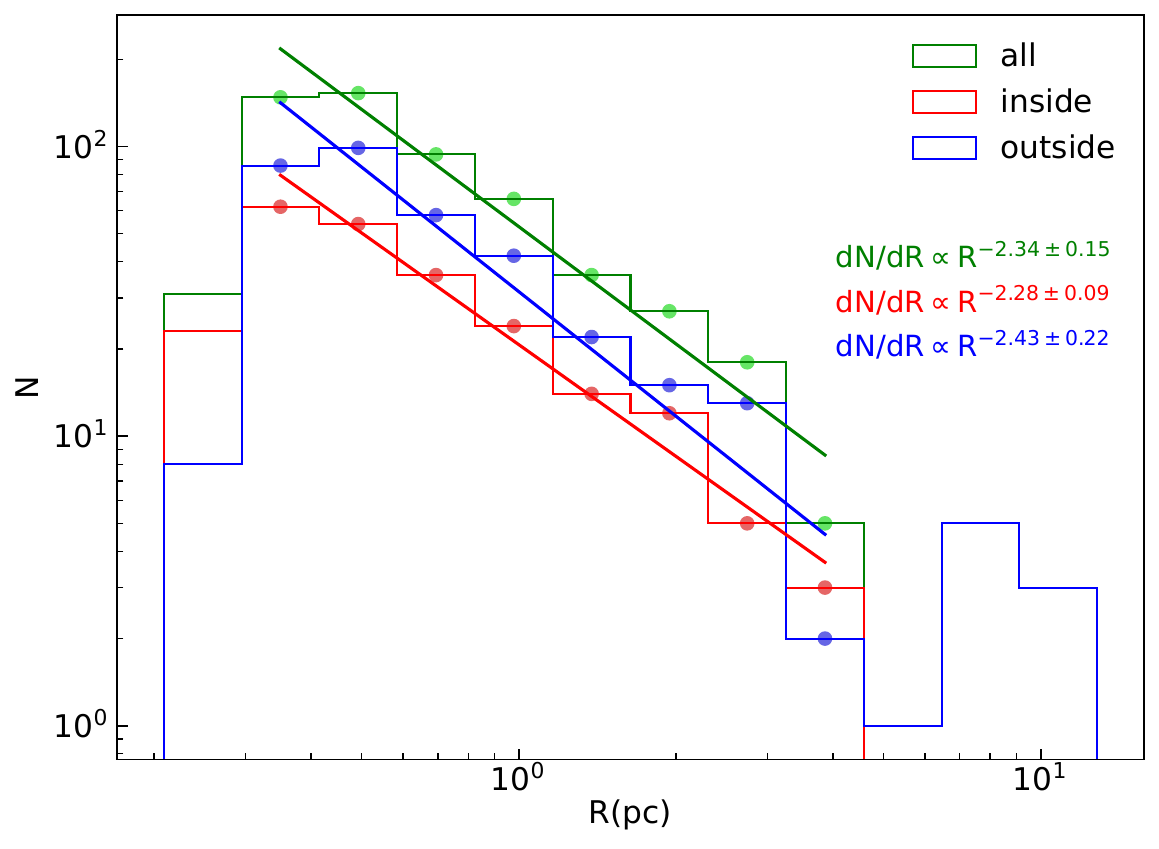}{0.5\textwidth}{(a)}
          \fig{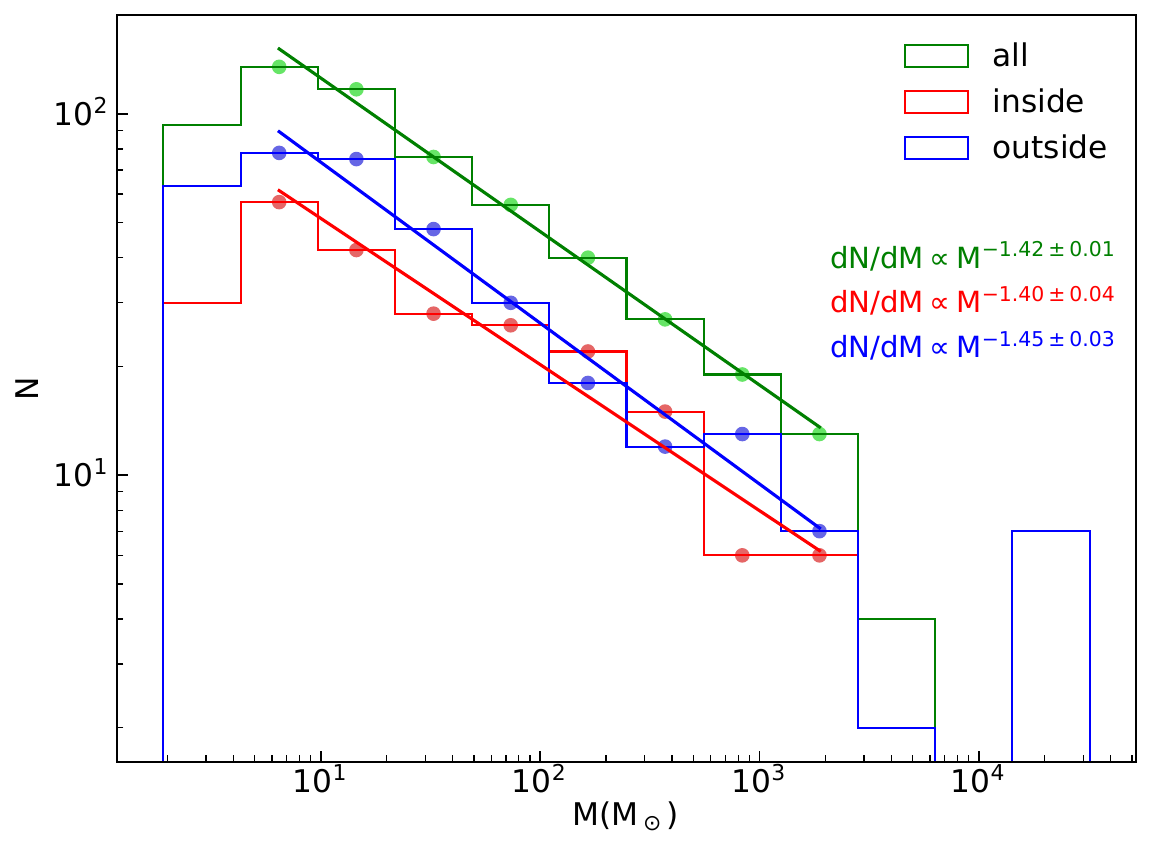}{0.5\textwidth}{(b)}}
	\gridline{
          \fig{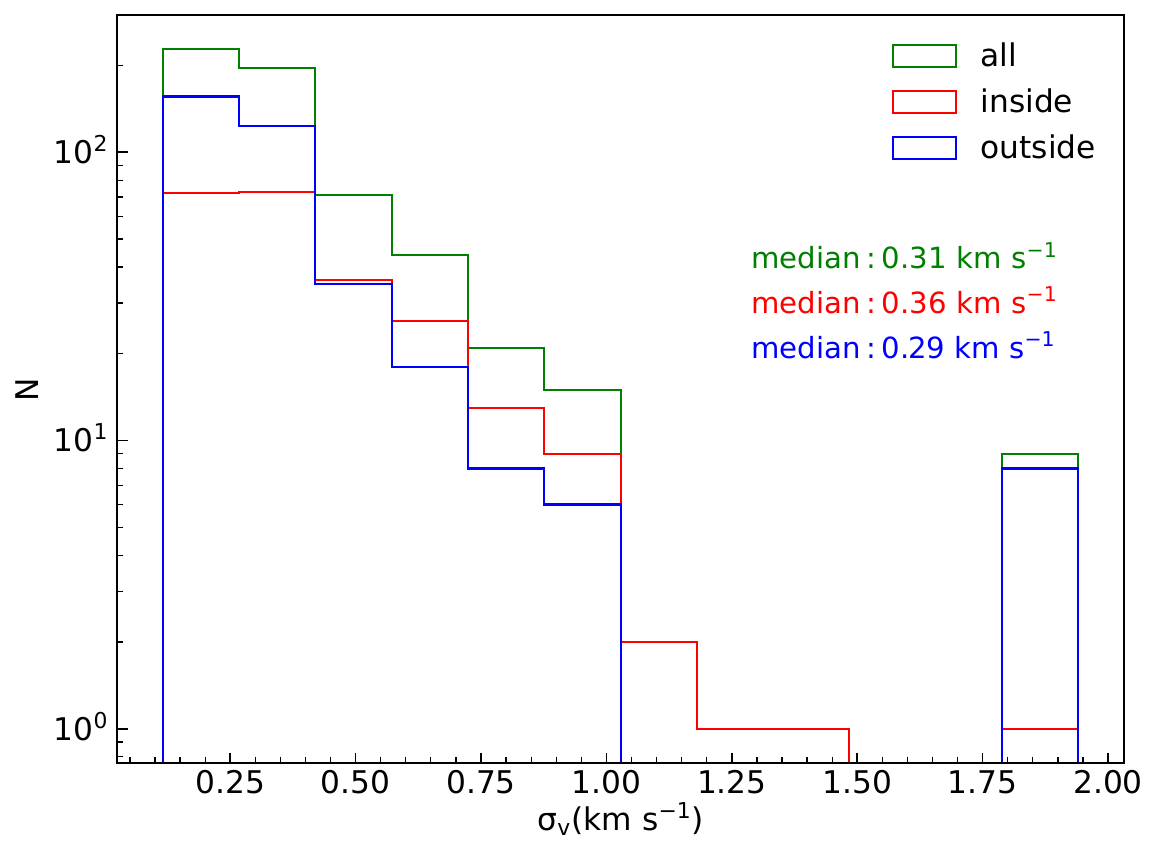}{0.5\textwidth}{(c)}
	 \fig{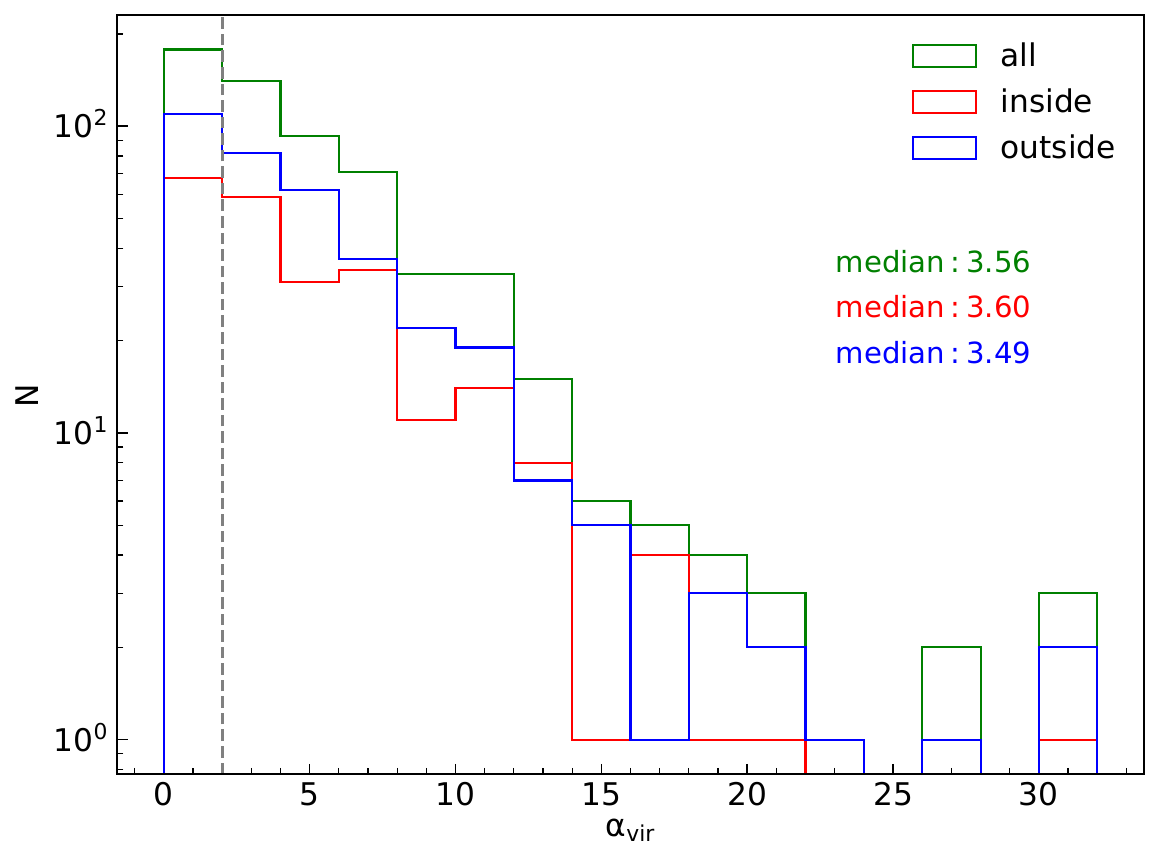}{0.5\textwidth}{(d)}
          }
    \caption{Histograms of the (a) radius, (b) mass, (c) velocity dispersion, and (d) virial parameter of the identified substructures. Colors in this figure have the same meanings as in Figure \ref{fig4}. The vertical grey dashed line in panel (d) presents $\alpha_{\text{vir}} = 2$.}
    \label{fig5}
\end{figure*}

In Figure \ref{fig5}, we present the statistical distributions of the mass, radius, velocity dispersion, and virial parameter for the identified substructures in the RMC. We fitted the mass and radius distributions with power-law functions based on the specific behaviors of these distributions. All the fitted power-law exponents are marked in Figure \ref{fig5}. We also divided these substructures into two groups based on whether they lie within or outside the \ion{H}{2} region. The groups inside and outside the \ion{H}{2} region are marked in red and blue, respectively, in Figure \ref{fig5}. 

The size distribution for the substructures is presented in Figure \ref{fig5}(a). For the entire RMC, the size distribution follows the relation $\Delta N/\Delta R\propto R^{-2.34\pm0.15}$, which is flatter than the power-law slope of $-3.2\pm0.1$ derived from the Outer Galaxy survey (OGS; \citealt{Heyer2001}), but comparable to the value of $-2.42 \pm 0.11$ reported by \citet{Ma2021} for molecular clouds in the Perseus arm identified using the MWISP data. The substructures inside and outside the \ion{H}{2} region exhibit consistent power-law radius distributions with that of the entire RMC. However, the structures inside the \ion{H}{2} region tend to follow a slightly shallower $R$ distribution than those outside the \ion{H}{2} region. The shallower radius distribution for substructures inside the \ion{H}{2} region indicate a higher fraction of larger structures than those outside the \ion{H}{2} region. For comparison, the radius distribution for substructures identified in the Maddalena GMC shows a slightly steeper power-law slope of -2.59 \citep{Shen2024}. 

In Figure \ref{fig5}(b), the mass distribution for substructures in the entire RMC yields a power-law relation of $\Delta N/\Delta M\propto M^{-1.42\pm0.01}$ within the range from $\sim$4 $M_{\odot}$ to 2$\times10^3\ M_{\odot}$. The mass distribution of substructures inside the \ion{H}{2} region follows $\Delta N/\Delta M \propto M^{-1.40\pm0.04}$, while that outside follows $\Delta N/\Delta M \propto M^{-1.45\pm0.03}$. These two distributions are consistent within statistical uncertainties. The fitted mass function is moderately steeper than the clump mass spectrum of the RMC reported by \citet{Williams1995}, who obtained a power-law index of $-1.27$, but shallower than the indices of $-1.61$ and $-1.80$ derived by \citet{Schneider1998} from $\rm ^{13}CO\ J=2-1$ and by \citet{Francesco2010} from \textit{Herschel} continuum data, respectively. Using $\rm ^{12}CO$ and $\rm ^{13}CO\ J=1-0$ observations from the FCRAO 14 m telescope together with \textit{Herschel} dust emission, \citet{Veltchev2018} obtained significantly steeper clump mass spectra, with indices ranging from $-2.1$ to $-3.3$. When comparing these studies, it is important to note that the previously reported distributions were derived specifically for core or clump structures, whereas our analysis is not restricted to a single hierarchical level, but covers a broad mass range, from below $10\ M_\odot$ to above $10^3\ M_\odot$. In addition, different structure identification methodologies can introduce systematic biases, as already shown by \citet{Veltchev2018}. Furthermore, the observed area in this work covers the entire molecular cloud (including both diffuse and dense gas), while that in \citet{Veltchev2018} is clump‑dominated. Systematic discrepancies also arise from tracer choice. As summarized in Table 5 of \citet{Veltchev2018}, dust-derived clump mass functions tend to show significantly steeper slopes than those obtained from CO data. In comparison, based on the MWISP data and using the same substructure identification algorithm as this work, \cite{Shen2024} reported a power-law index of -1.64 for the Maddalena GMC in the mass range from 10 $M_\odot$ to $1\times 10^3\ M_\odot$. We can see that the fraction of relatively massive substructures is higher in the RMC than the Maddalena GMC. 

Figure \ref{fig5}(c) shows the distribution of the velocity dispersion $\sigma_v$, defined as the intensity-weighted second moment of the velocities within the PPV volume of each structure. The median velocity dispersions are 0.31, 0.36, and 0.29 $\rm km \ s^{-1}$ for structures in the entire RMC, within the \ion{H}{2} region, and outside the \ion{H}{2} region, respectively, which agree well with the median velocity dispersion of $\rm ^{13}CO$ substructures in the Maddalena GMC with the MWISP data and using Dendrogram \citep{Shen2024}. Although the median values of $\sigma_v$ for structures inside and outside the \ion{H}{2} region are similar, there appear to be more substructures with $\sigma_v$ greater than 0.5 km s$^{-1}$ within the \ion{H}{2} region. Based on the distribution of $\sigma_v$ and the measured median values, we can roughly estimate the range and typical value of the sonic Mach number, using $M_s = \sqrt{3} \sigma_{nt}/c_s$, where $\sigma_{nt}$ is the one-dimensional non-thermal velocity dispersion, defined as $\sigma_{nt} = \sqrt{\sigma_v^2 - \sigma_{thermal}^2}$, and $c_s$ is the sound speed. Assuming a kinetic temperature of 10 K, the sound speed is $c_s = \sqrt{k_B T_k/(\mu_Hm_H)} \approx 0.187\sqrt{T_{\rm kin}/10} \approx 0.2$ km s$^{-1}$ \citep{Schneider2013}, where $k_{B}$ is the Boltzmann constant, $\mu_{H}=2.37$ is the mean molecular weight per free particle \citep{Kauffmann2008}, and $m_{H}$ is the mass of atomic hydrogen. The Mach numbers of all substructures lie in the range from 1 to 17, with a typical value of around 3. 

Figure \ref{fig5}(d) presents the distribution of the virial parameter. The $\rm ^{13}CO$ structures in the RMC have a median $\alpha_{\rm vir}$ of 3.6, and 30\% are subvirial ($\alpha_{\text{vir}} < 2$). The fraction of gravitationally bound structures is similar for inside and outside \ion{H}{2} region, which have median $\alpha_{\text{vir}}$ values of 3.56 and 3.49, respectively. Therefore, although the substructures inside and outside the \ion{H}{2} region exhibit modest differences in mass, radius, and hierarchical complexity, they are similar in terms of gravitational binding state.

\subsection{Scaling Relations}\label{sec3.3}

\begin{figure*}[!htb]
	\gridline{\fig{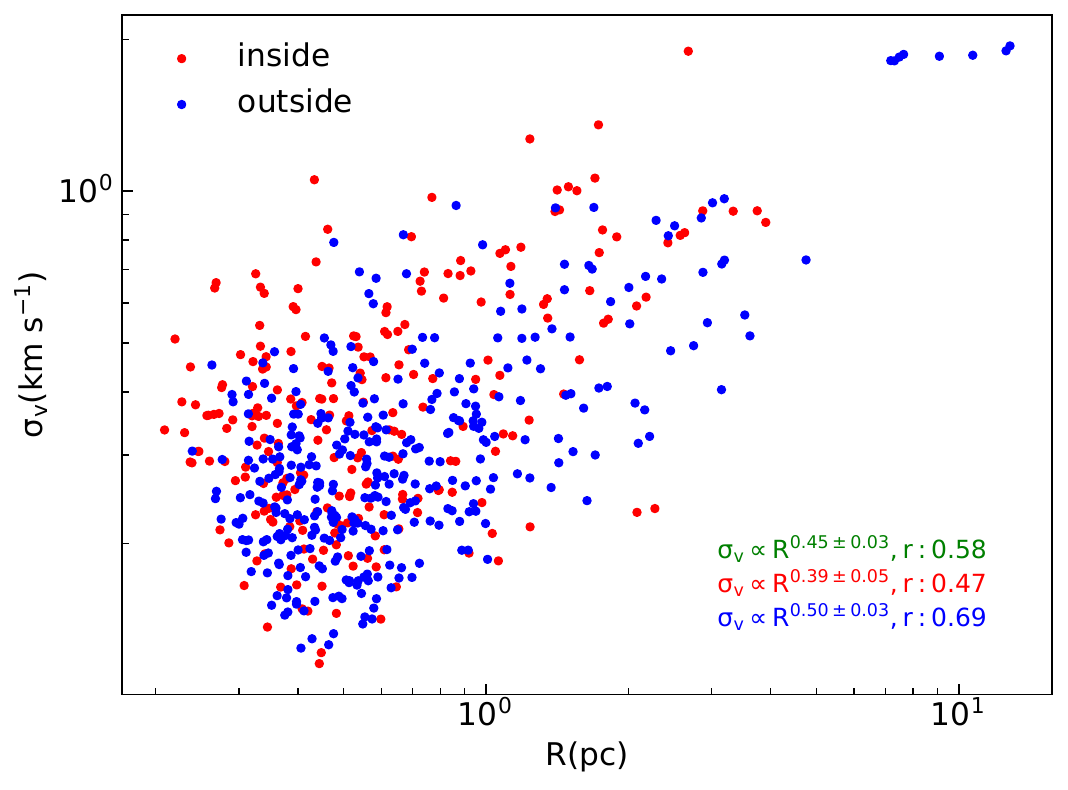}{0.5\textwidth}{(a)}
	\fig{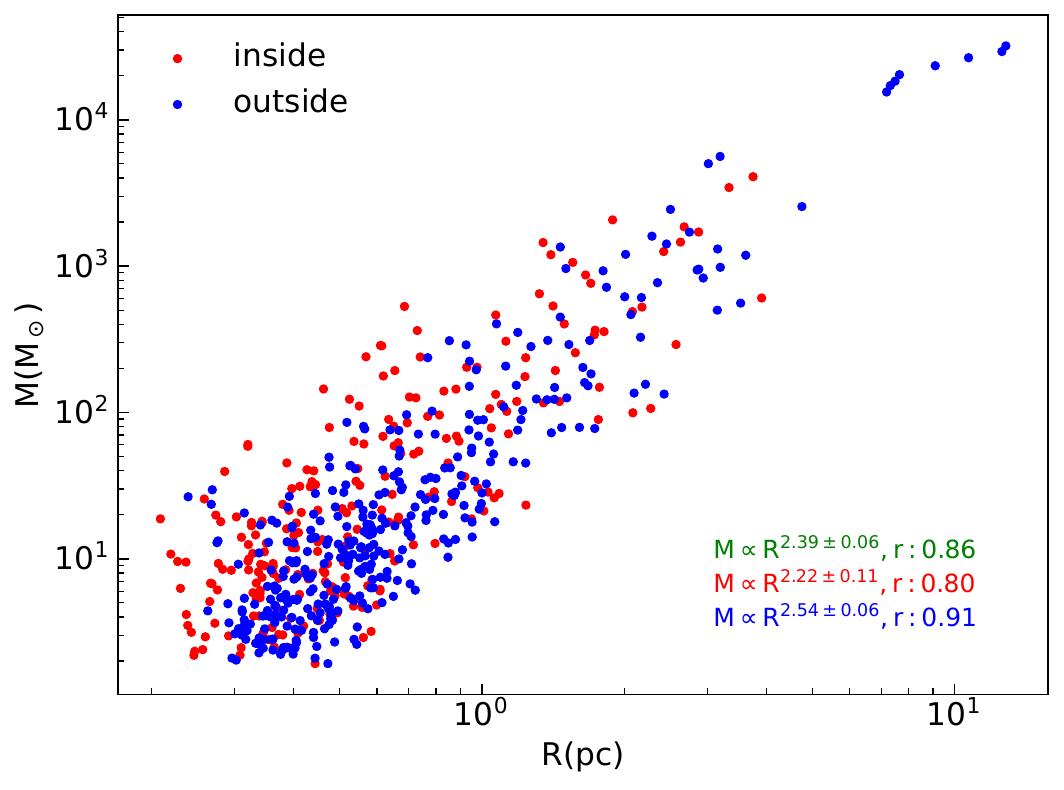}{0.5\textwidth}{(b)}}
	\gridline{
	\fig{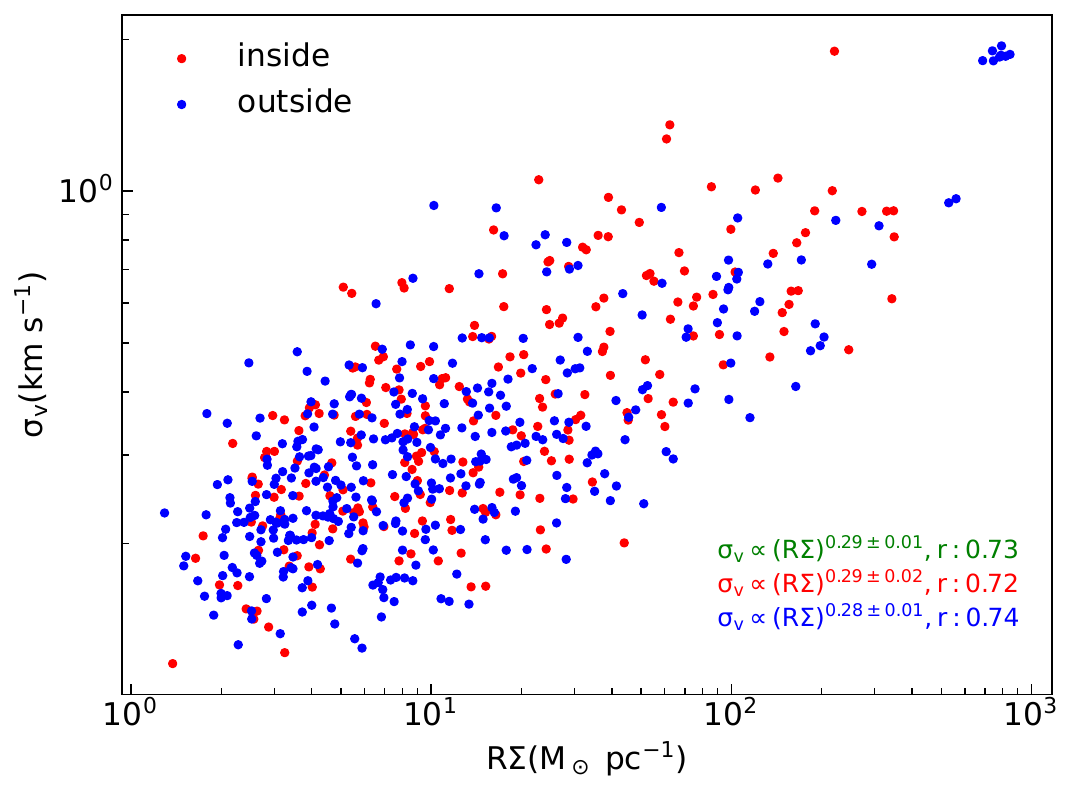}{0.5\textwidth}{(c)}
	\fig{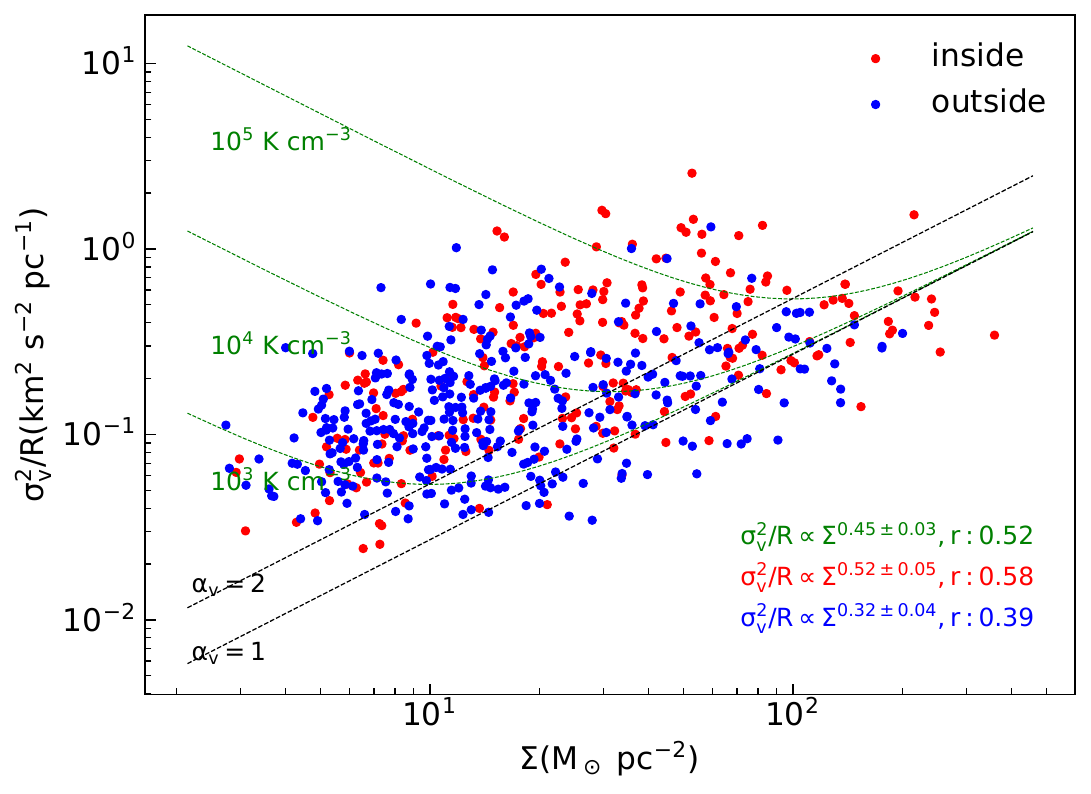}{0.5\textwidth}{(d)}
	}
\caption{Relationships between (a) $\sigma_v$ and $R$, (b) $M$ and $R$, (c) $\sigma_v$ and $R\Sigma$, and (d) $\sigma_v^2/R$ and $\Sigma$ of the identified substructures in the RMC. The two black dashed lines in panel (d) represent $\alpha_{\text{vir}}$ equals 1 (simple virial equilibrium) and 2 (marginal equilibrium), respectively. The green dashed lines indicate the solutions for the pressure-confined virial equilibrium state for varying external pressures. In all the panels, the red dots represent the structures inside the \ion{H}{2} region while the blue ones indicate those outside the \ion{H}{2} region. Text in the bottom right corner indicates the fitting results and the corresponding Pearson correlation coefficients, with green, blue, and red colors representing the fits for all substructures, substructures outside \ion{H}{2} regions, and substructures inside \ion{H}{2} regions, respectively.}\label{fig6}
\end{figure*}

In this section, we present the analyses of the scaling relations between the measured physical parameters of the structures. Figure \ref{fig6}(a) shows the $\sigma_v-R$ relationship of the structures identified in this work, with different colors indicating their locations, i.e., whether they lie within the \ion{H}{2} region. The $\sigma_v-R$ relation for all the structures identified in this work is $\sigma_v \propto R^{0.45\pm0.03}$, and the Pearson correlation coefficient between $\log {\sigma_v}$ and $\log R$ is 0.58. However, the $\sigma_v-R$ relations for the structures inside and outside the \ion{H}{2} region have moderately different exponents: $0.39\pm0.05$ and $0.50\pm0.03$, respectively. The Pearson correlation coefficient is 0.69 for the structures outside the \ion{H}{2} region, which is higher than the value of 0.47 for those inside the \ion{H}{2} region. The power-law indices obtained in this work lie between the index of the original Larson's relation (0.38) and the widely accepted value of 0.5 for Burgers' turbulence \citep{Solomon1987, Heyer2004, Rice2016, Hacar2023}. Using clumps derived from the RMC, \citet{Veltchev2018} obtained a near-flat linewidth-size relation, with the velocity dispersions of all the clumps appear to be only slightly suprathermal. This discrepancy likely stems from differences in both the structures of interest and the structure-identification algorithms employed, as the density threshold used in their analysis is significantly higher.

Figure \ref{fig6}(b) illustrates the correlation between the mass and radius of the structures. It reveals a relation of $M\propto R^{2.39\pm0.06}$ for all the structures, with a power-law index similar to that obtained with previous Galactic molecular cloud catalogs (e.g., $2.2\pm0.2$ from \citealt{Miville2017}). Using $\rm ^{12}CO$, $\rm ^{13}CO\ J=1-0$, and dust emission, \cite{Veltchev2018} obtained similar mass-size relation for clumps in the RMC ($\rm 2.5\pm 0.3, 2.6\pm 0.5, and\ 2.3\pm0.5$ for clumps identified with $\rm ^{12}CO$, $\rm ^{13}CO$, and dust emission, respectively). The power-law index derived for structures outside the \ion{H}{2} region is $2.54\pm0.11$, slightly higher than that for structures inside ($2.22\pm0.06$), suggesting that mass increases more rapidly with radius in the outside region. At a given radius, substructures inside the \ion{H}{2} region tend to have higher masses, indicating higher surface densities.

\begin{figure*}[!htb]
    \centering
    \includegraphics[width=\textwidth]{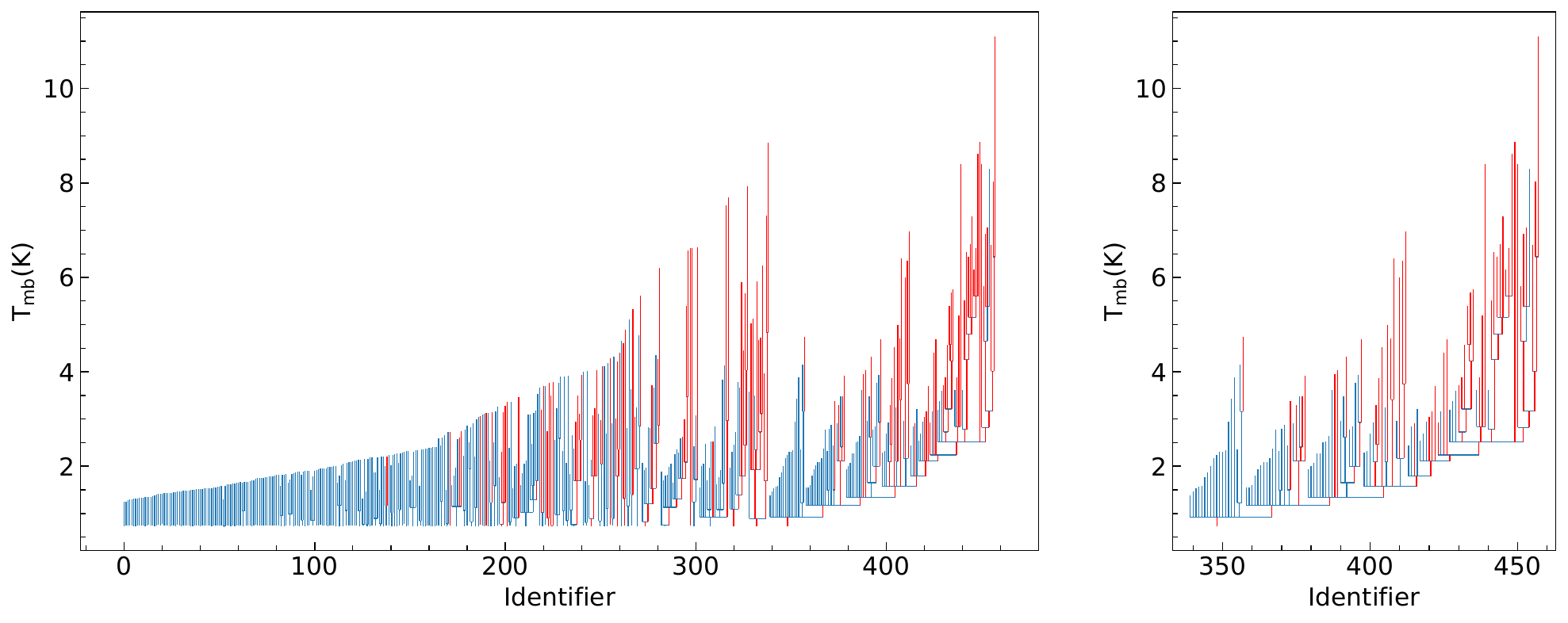}
    \caption{Left: the tree diagram of substructures identified by the modified non-binary Dendrogram algorithm. Right: a zoom-in of the biggest tree in the left panel. Red color marks the gravitationally bound structures while blue color marks those unbound.}\label{fig7}
\end{figure*}

Figure \ref{fig6}(c) presents the relation between $\sigma_v$ and $\Sigma R$. When considering the surface densities of the structures, we find the relation $\sigma_v\propto(\Sigma R)^{0.29\pm0.01}$. The $\sigma_v-(\Sigma R)$ relation shows a stronger correlation than the $\sigma_v-R$ relation, with a Pearson correlation coefficient of 0.73 for $\log {\sigma_v}$ and $\log {(\Sigma R)}$. No significant difference is found between structures inside and outside the \ion{H}{2} region in the $\sigma_v\propto (\Sigma R)$ relation. The exponent of the $\sigma_v-(\Sigma R)$ relation we obtained differs significantly from the previously predicted and observed value of 0.5 \citep{Heyer2009}. This discrepancy might suggest a different physical origin for the $\sigma_v-(\Sigma R)$ relation, rather than due to self-gravity.

Another physical interpretation of the underlying relationship between the observed line-width and size of the molecular clouds considers the external pressure in the equilibrium of the clouds. The solution for balance between external pressure, internal pressure, and self-gravity can be delineated by ``V'' shaped lines in the $\sigma_v^2/R-\Sigma$ parameter space. Figures \ref{fig6}(d) presents the observed $\sigma_v^2/R-\Sigma$ relation of substructures in the RMC. These substructures show neither preferential distribution along the lines where the virial parameter equals 1 or 2, nor clustering around a characteristic external pressure, suggesting that either pressure plays a relatively minor role, or its value varies significantly across different structures. The substructures exhibit slightly different trends when considering their locations with respect to the \ion{H}{2} region. We fitted the $\sigma_v^2/R$-$\Sigma$ relation with a power-law function for all the substructures, as well as for those inside and outside the \ion{H}{2} region, respectively. The best-fit relations are $\sigma_v^2/R \propto \Sigma^{0.45 \pm 0.03}$ for all substructures, $\sigma_v^2/R \propto \Sigma^{0.52 \pm 0.05}$ for those inside, and $\sigma_v^2/R \propto \Sigma^{0.32 \pm 0.04}$ for those outside the \ion{H}{2} region. The power-law index closer to 0.5 for substructures inside the \ion{H}{2} region may suggest that gravity plays a more significant role in influencing these structures compared to those outside the \ion{H}{2} region.
We conducted test runs on the obtained scaling relations against different sets of input parameters, specifically $min\_delta$ and $min\_value$, for the non-binary Dendrogram algorithm. The results of these test runs are presented in Section \ref{secB1} in the Appendix. The results observed in this section remain robust against variations in the input parameters.

\section{Discussion}\label{sec4}
\subsection{Which Part of the RMC is Gravitationally Bound?}

\begin{figure}[!htb]
	\fig{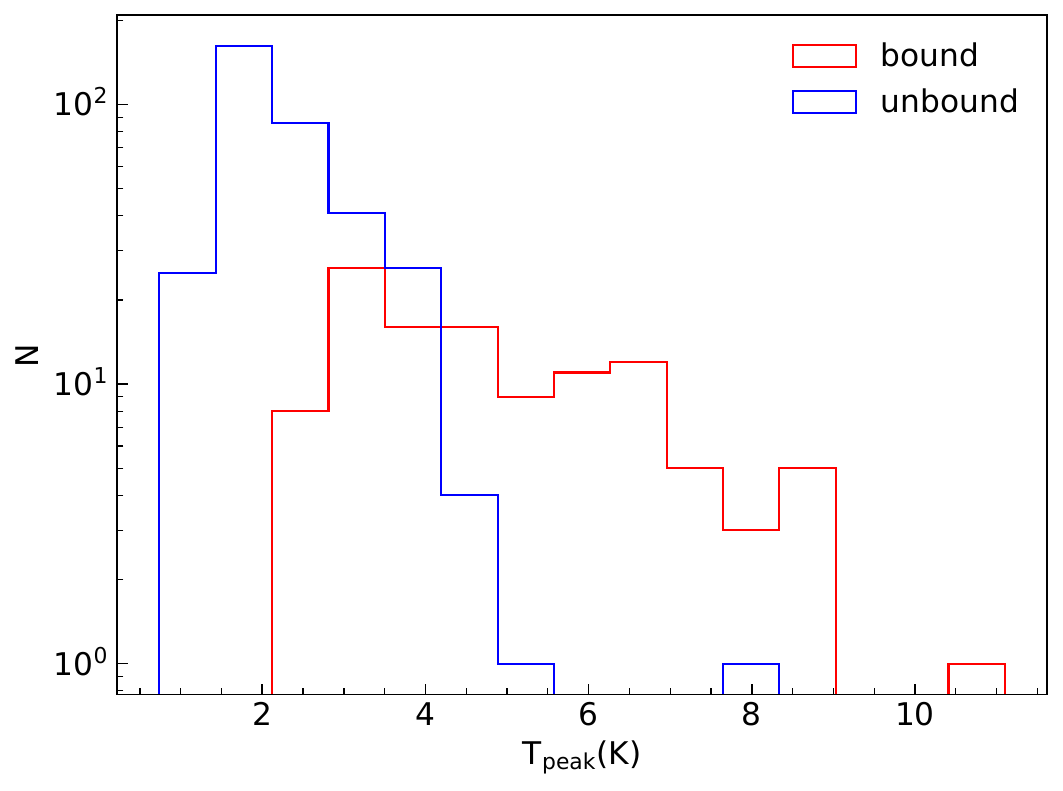}{0.4\textwidth}{}
\caption{Histogram of peak brightness temperature of the bound (red) and unbound (blue) leaves in the tree shown in Figure \ref{fig7}.}\label{fig8} 
\end{figure}

\begin{figure*}[!htb]
	\gridline{\fig{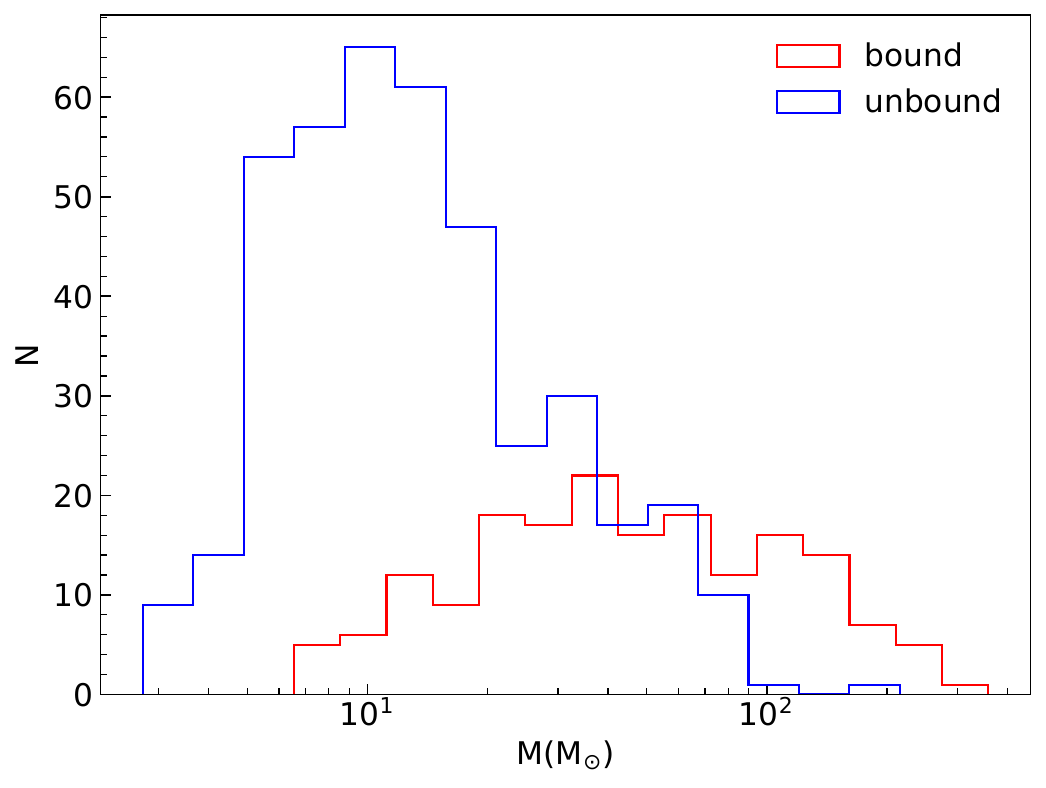}{0.5\textwidth}{(a)}
	\fig{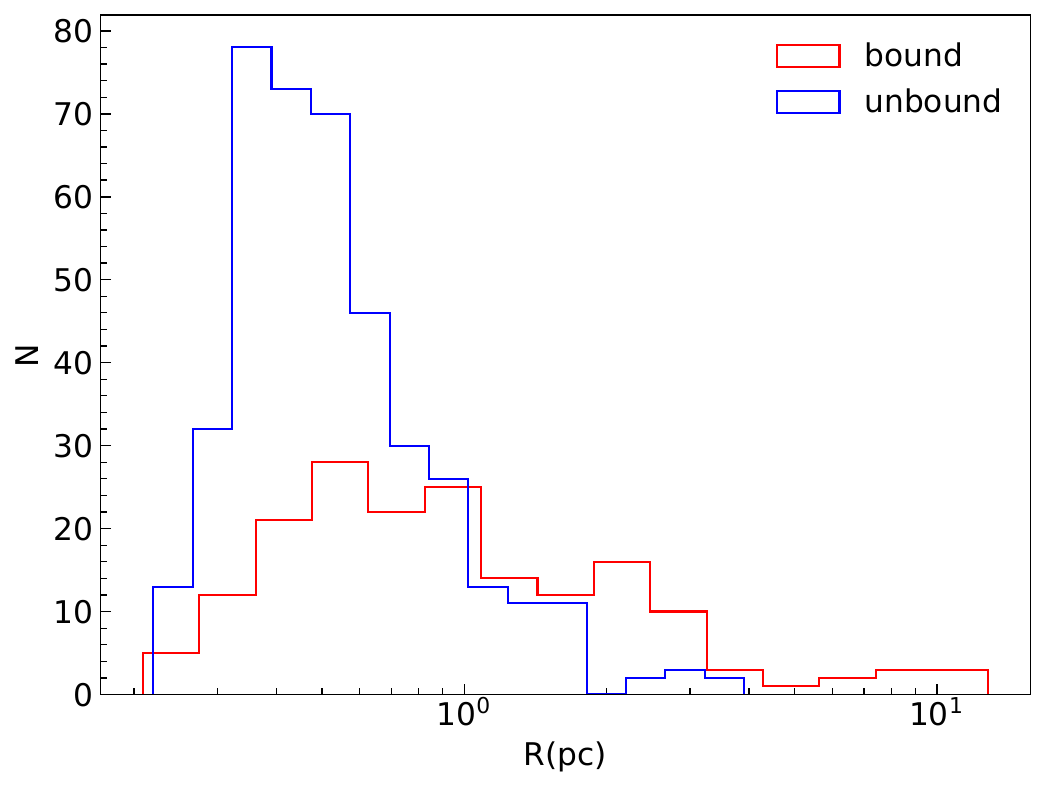}{0.5\textwidth}{(b)}}
	\gridline{\fig{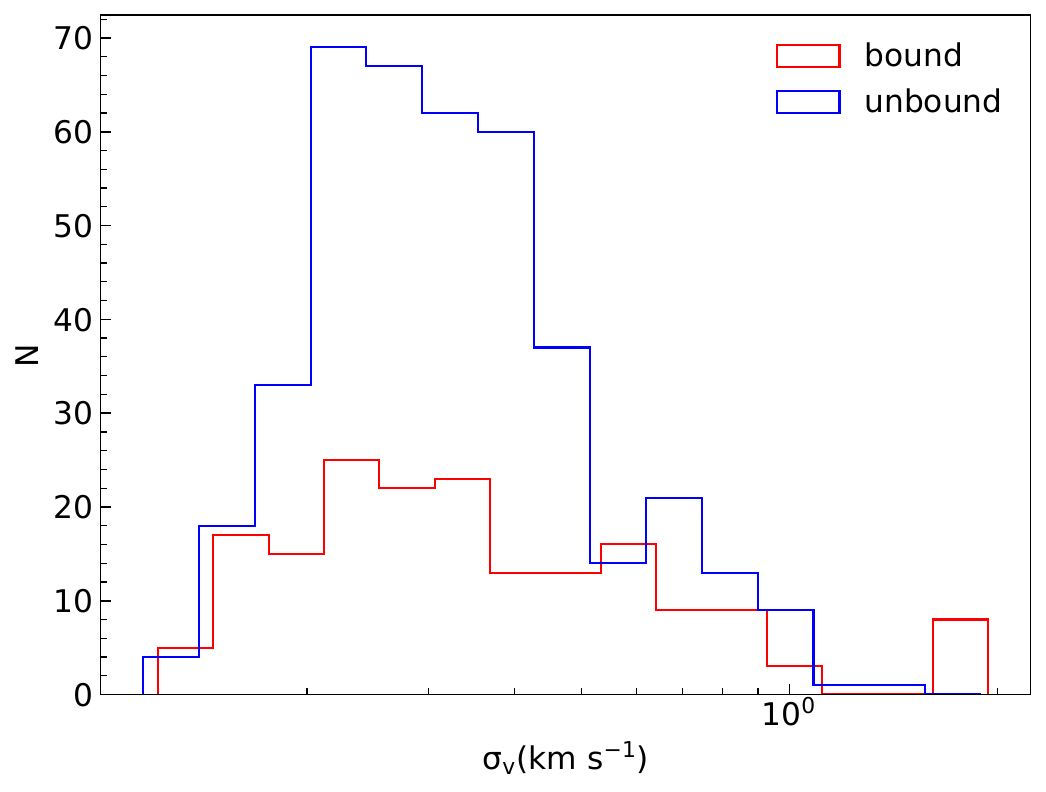}{0.5\textwidth}{(c)}
	\fig{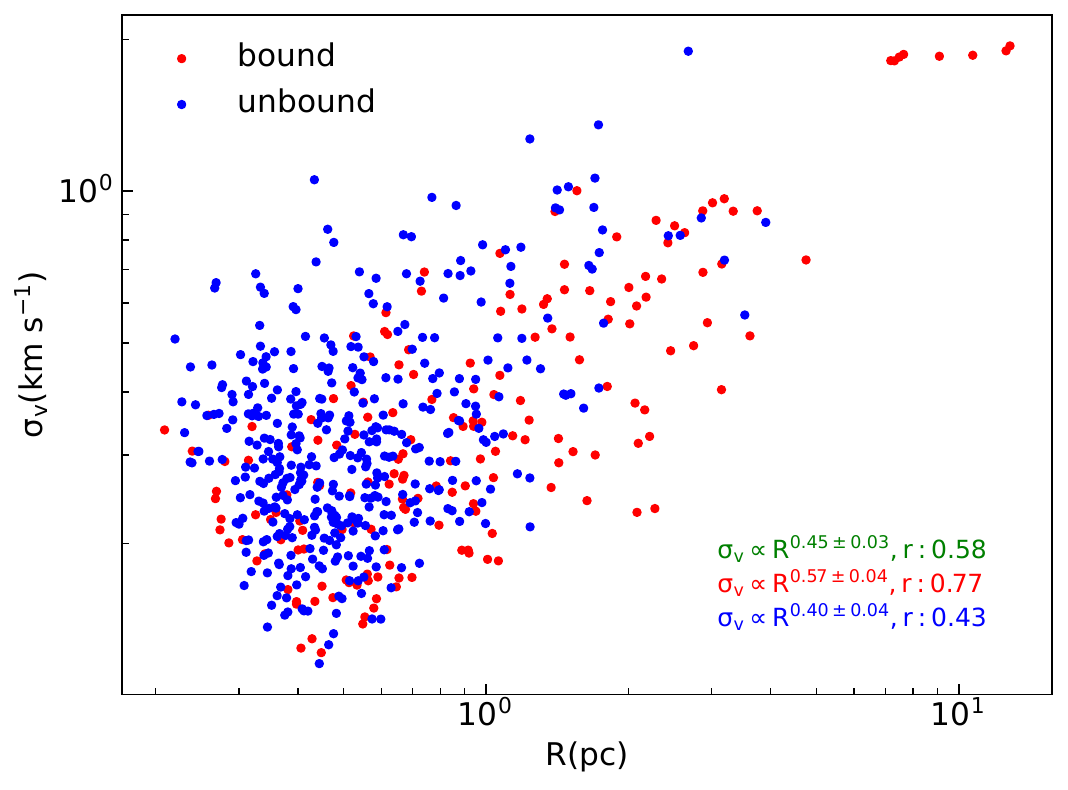}{0.5\textwidth}{(d)}}
\caption{Histograms of the (a) mass, (b) radius, and (c) velocity dispersion of the bound (red) and unbound (blue) substructures in the tree shown in Figure \ref{fig7}. (d) Relationships between radius and velocity dispersion for the bound (red) and unbound (blue) substructures. The fitting results are presented in text with corresponding colors, while the green one represents the fitting result for all substructures.}\label{fig9} 
\end{figure*}

In Figure \ref{fig7}, the gravitationally bound structures are highlighted in red in the tree diagram. For clarity, the right panel of Figure \ref{fig7} provides a zoomed-in view of the biggest tree from the left panel. The figure shows that the gravitationally bound leaves predominantly reside in the upper parts of the trunks. Across all substructures shown in the right panel of Figure \ref{fig7}, all but five self-gravitating leaves are contained within self-gravitating parental structures. These results are consistent with those obtained by \cite{Goodman2009} in the L1448 star-formation region within the spatial range from $\rm \sim 0.1 \ pc$ to 1 pc, and we extend the spatial scale of gravitationally bound structures to $\sim$10 pc in the RMC. However, some different results can be seen between the two works. In Figure 2 of \cite{Goodman2009}, although self-gravitating branches exist on all possible spatial scales, only a few of the leaves are self-gravitating. In this work, nearly all leaves above $\sim$4 K are self-gravitating, as revealed in Figure \ref{fig8}. Figure \ref{fig8} shows that the self-gravitating leaves possess significantly greater $T_{\text{peak}}$ than the unbound leaves. Consistently, three of the four self-gravitating leaves in \cite{Goodman2009} are found in regions above $\sim$ 4 K.

Figure \ref{fig9}(a) shows the distribution of mass of the substructures in the RMC, grouped by their gravitational binding states. Gravitationally bound structures, shown in red, tend to have systematically higher masses. The most massive structures (up to $\> 2 \times 10^3 M_{\odot}$) are all gravitationally bound, while low-mass structures (e.g., $<$10 $M_{\odot}$) are most likely to be gravitationally unbound. This trend is expected, as an anti-correlation between mass and virial parameter is widely observed (e.g., \citealp{Kauffmann2008} and references therein).

Figure \ref{fig9}(b) illustrates the radius histogram of structures in different gravitational binding states. The gravitationally bound structures, colored in red, display a broad range in radius, whereas unbound structures are mostly concentrated within 2 pc. And Figure \ref{fig9}(b) suggests that substructures with radii larger than 2 pc are most likely bound. 

Figure \ref{fig9}(c) shows the velocity dispersion histogram of the substructures. Unlike the distributions of mass and radius, the range of velocity dispersion shows negligible difference between structures in different gravitational binding states. However, considering that gravitationally bound structures exhibit relatively larger scales and masses, as shown in panels (b) and (c), the comparable velocity dispersions suggest that turbulence is relatively weaker within these structures. 

Figure \ref{fig9}(d) presents the correlation between radius and velocity dispersion, similar to Figure \ref{fig6}(a), but with structures highlighted by their gravitationally binding states. The turbulent support model and the GHC scenario provide different interpretations of the observed linewidths. The former attributes the velocity dispersion of molecular clouds to the scale-free nature of supersonic turbulence (e.g., \citealp{Kritsuk2013, Padoan2016}), whereas the latter argues that a large fraction of the velocity dispersion originates from gravitational collapse, with the turbulence itself being only transonic. The linewidth–size relation alone is insufficient to distinguish between the proposed scenarios. Additional observational constraints are needed to reach a conclusive interpretation. Nevertheless, it is evident that unbound structures tend to exhibit slightly stronger turbulent motions than gravitationally bound ones. 


\begin{figure*}[!htb]
	\gridline{\fig{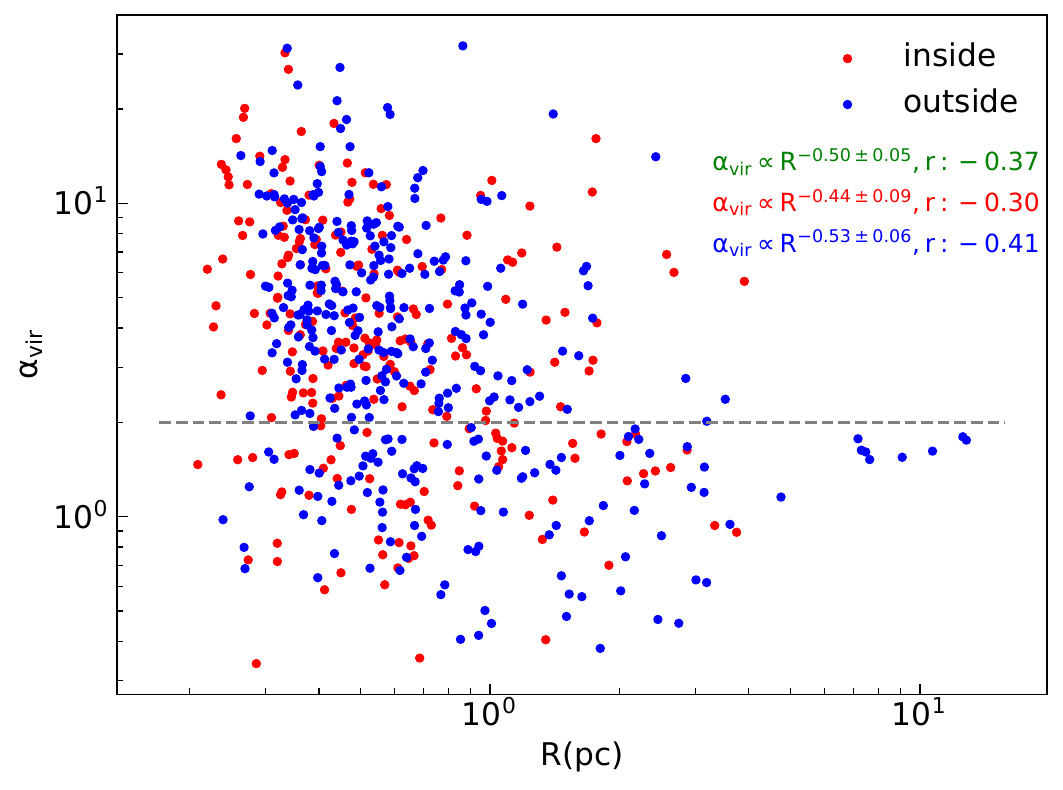}{0.4\textwidth}{(a)}
	\fig{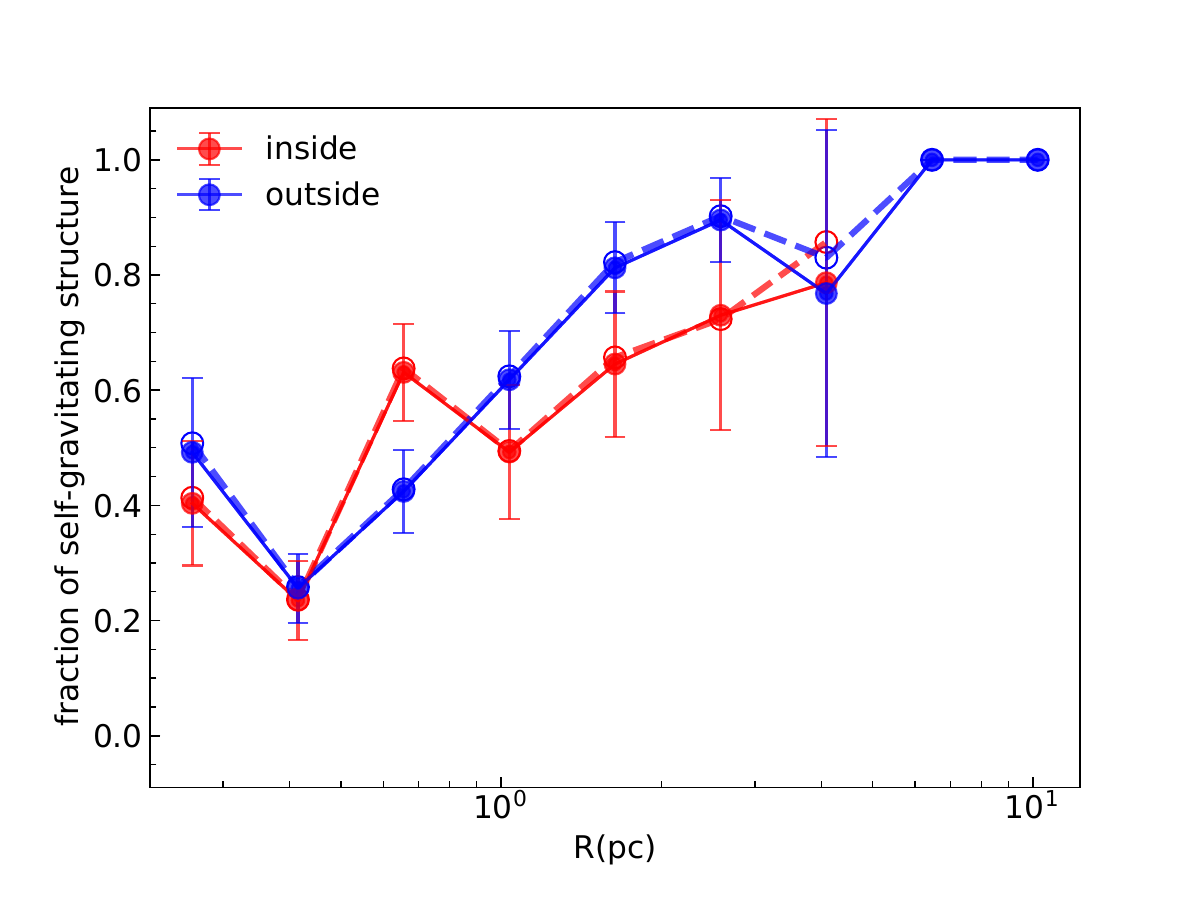}{0.46\textwidth}{(b)}}
	\vspace{-0.7cm}
	\gridline{\fig{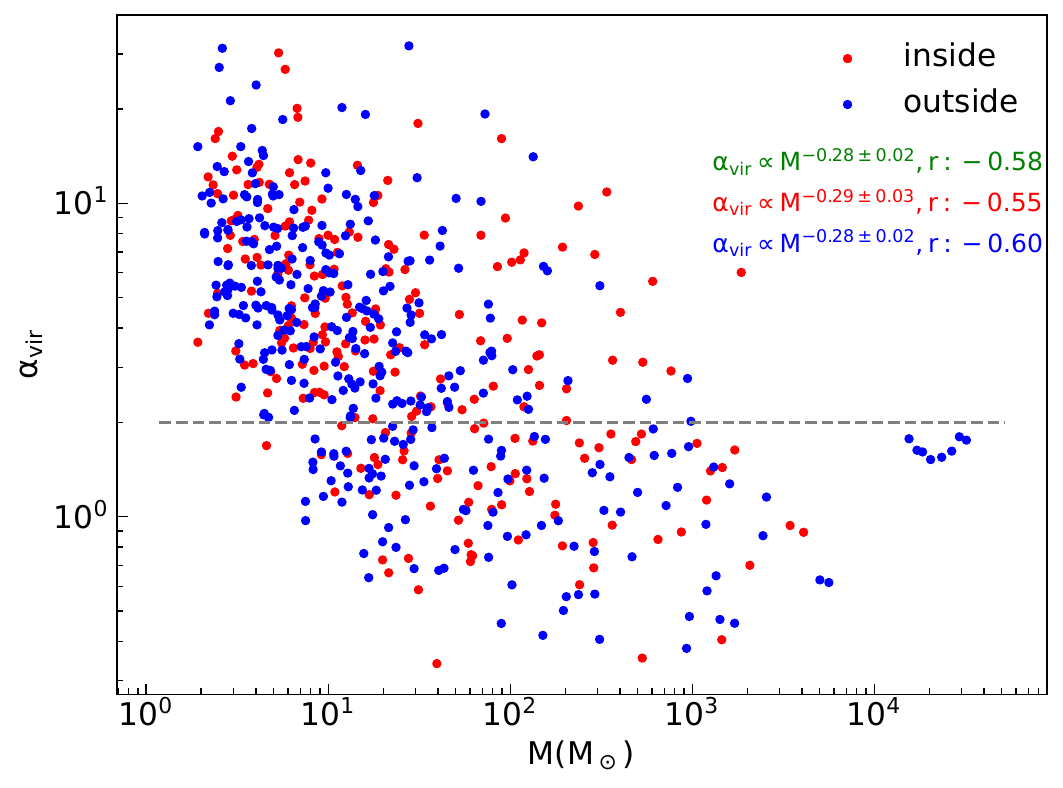}{0.4\textwidth}{(c)}
	\fig{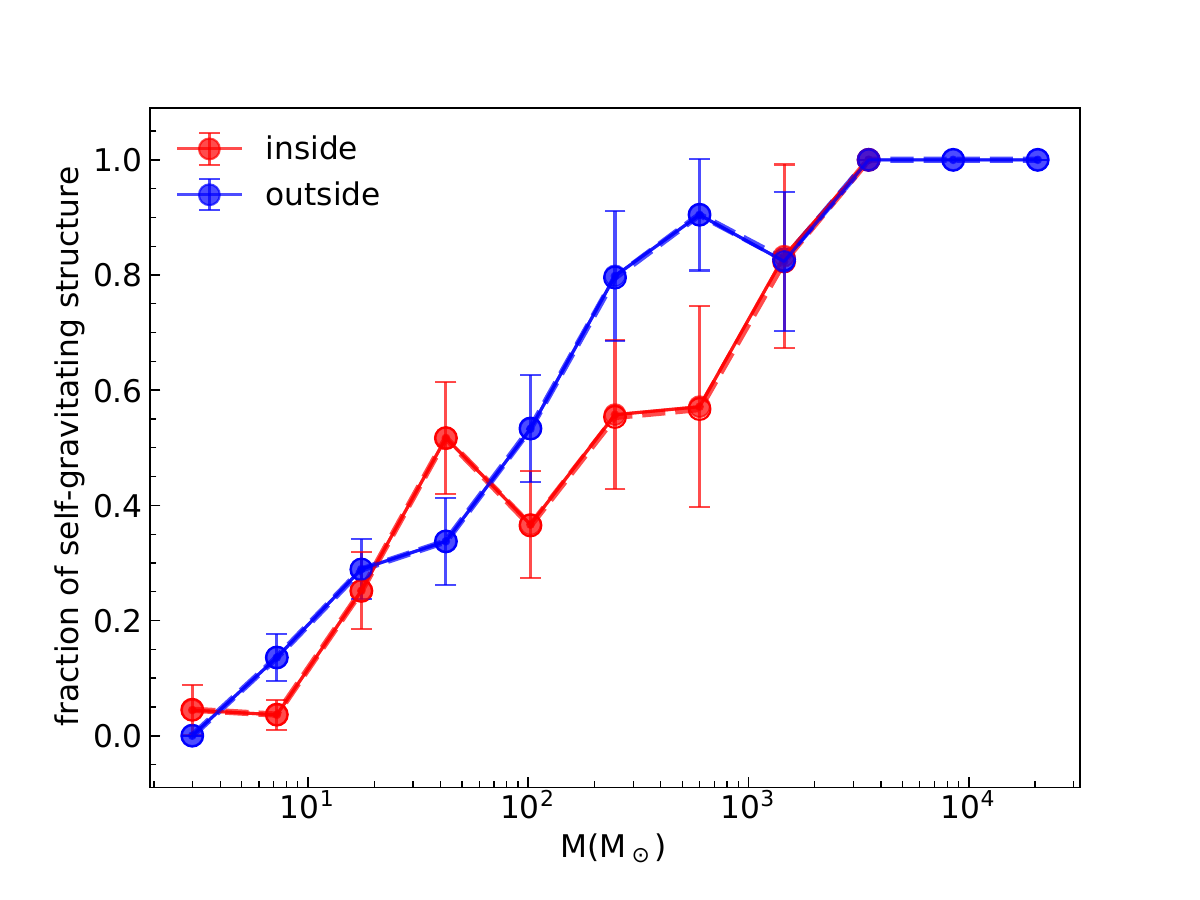}{0.46\textwidth}{(d)}}
	\vspace{-0.7cm}
	\gridline{\fig{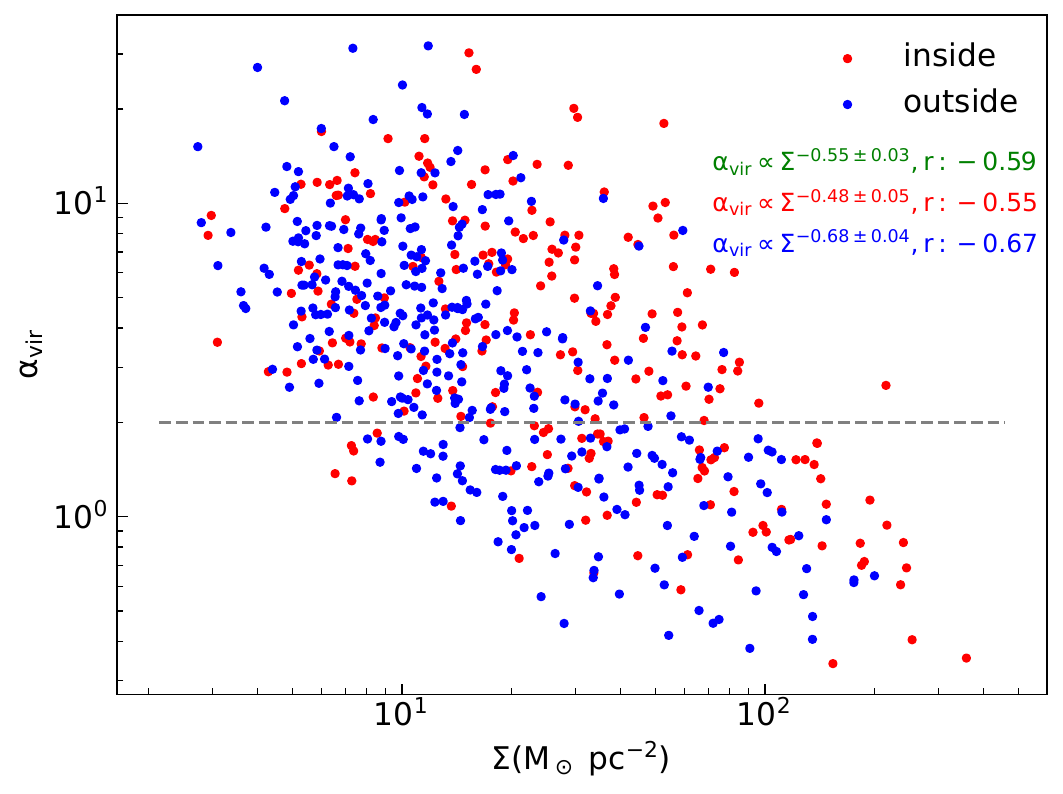}{0.4\textwidth}{(e)}
	\fig{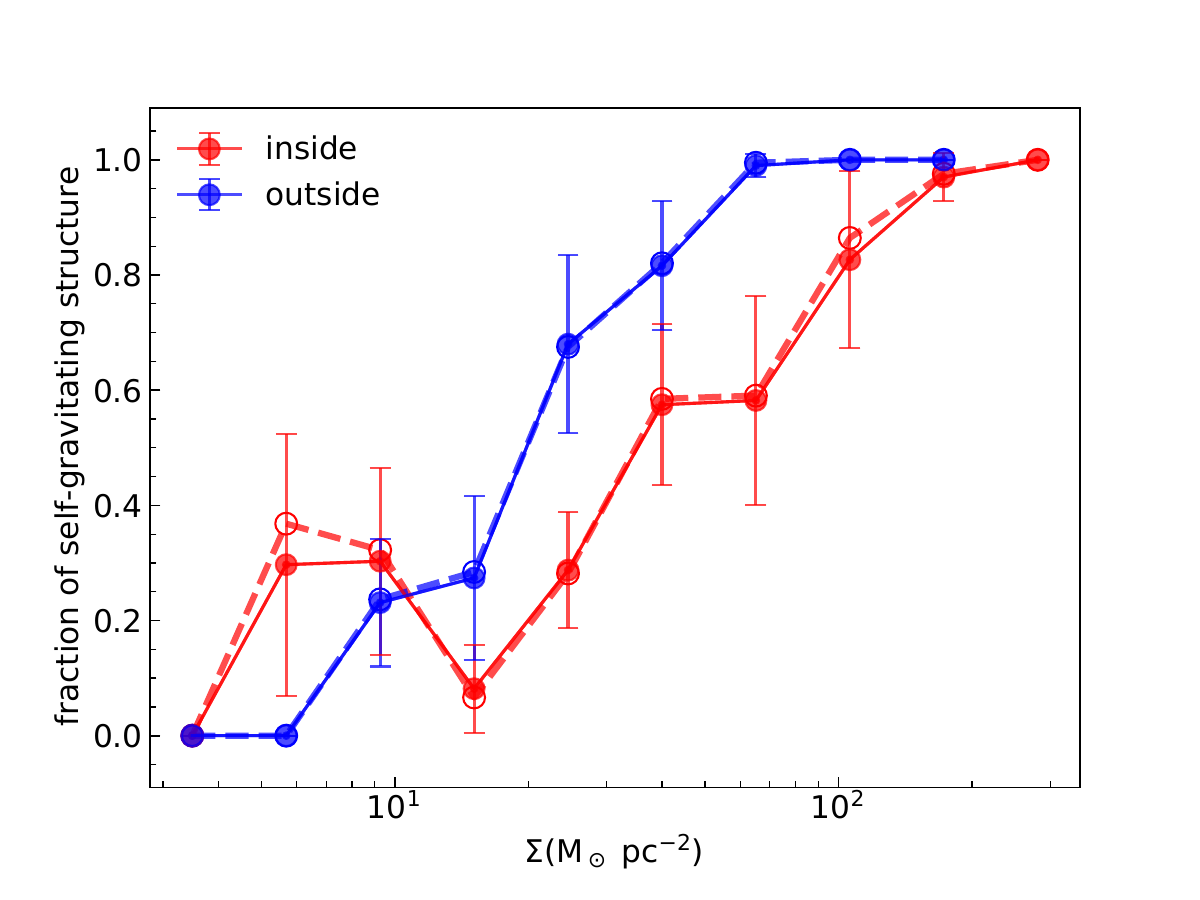}{0.46\textwidth}{(f)}}
\caption{(a) Correlation between virial parameter and radius of the substructures in the RMC. (b) Fraction of gravitationally bound substructures as a function of the radius of the substructures. Filled symbols represent the averages from the bootstrap resampling and are connected with solid lines, while hollow symbols show the results without bootstrap resampling, and are connected with dashed lines. (c) Same as panel (a) but for the relation between $\alpha_{\text{vir}}$ and $M$. (d) Same as panel (b) but for the variation against mass. (e) Same as panel (a) but for the relation between $\alpha_{\text{vir}}$ and surface density. (f) Same as panel (b) bur for the variation against surface density. The grey horizontal dashed lines in panels (a) and (c) highlight the critical virial parameter of $\alpha_{\text{vir}}=2$. The red color and blue color in the four panels indicate structures inside and outside the \ion{H}{2} region, respectively. The text in panels (a), (c), and (e) indicates the corresponding fitting results, where the green text represents the fitting results for all substructures.}\label{fig10}
\end{figure*}

\subsection{Variation of Equilibrium States with Radius, Mass, and Surface Density}\label{sec4.2}

To compare with the results from \cite{Goodman2009}, we examine how the virial parameter varies with radius, mass, and surface density. Figure \ref{fig10}(a), (c), and (e) show the $\alpha_{\text{vir}}-R$, $\alpha_{\text{vir}}-M$, and  $\alpha_{\text{vir}}-\Sigma$ relations for the substructures, respectively. The grey dashed lines in the three panels represent $\alpha_{\text{vir}} = 2$, below which structures are considered to be dominated by gravity. The virial parameter systematically decreases with radius, mass, and surface density, though considerable scatter exists. \citet{Veltchev2018} found that clumps in the Rosette Molecular Cloud follow the relation $\alpha_{\text{vir}} \propto M^{-\epsilon}$ with $\epsilon \sim 0.6$–$0.7$, while we obtain a value around 0.3 for the dendrogram structures. This discrepancy may be related to the fact that the two algorithms identify structures of different physical nature. Veltchev et al. identified dense clumps with relatively high intensity thresholds, whereas our dendrogram approach recovers a hierarchical set of structures that simultaneously includes both diffuse, low-density components and compact, massive ones. The broader mass range and the different physical characteristics of the identified structures may therefore contribute to the shallower slope in our results.
Figure \ref{fig10}(b), (d), and (f) show the fraction of self-gravitating structures as a function of radius, mass, and surface density, respectively. The calculations for these relationships adopt the method in \cite{Shen2024}. The fraction represents the ratio of the luminosity of self-gravitating structures to the total luminosity within each radius, mass, or surface density bin. The error bars are derived from 10,000 bootstrap resampling iterations applied to the full sample of substructures.  In the three relationships, the proportion of self-gravitating structures increases gradually with radius, mass, and surface density, over the ranges of $\sim$0.25 pc to 10 pc, 3 M$_{\sun}$ to 3$\times$10$^{4}$ M$_{\sun}$, and $\rm \sim 4 \ M_\odot \ pc^{-2}$ to  $\rm 300 \ M_\odot \ pc^{-2}$, respectively. Minor decreases are observed at radius of 0.4 and 1 pc, mass of $\rm 100 \ M_\odot$, and surface density of $\rm 10 \ M_\odot \ pc^{-2}$ for substructures inside the \ion{H}{2} region, but the overall trend remains upward. For structures outside the \ion{H}{2} region, slightly higher fractions are found within the radius range of 1–4 pc, mass range of 100–2000 M${\sun}$, and surface density range of 10-200 $\rm M_\odot \ pc^{-2}$. Consistent behaviors are observed when considering only the largest tree in the RMC, as shown in Figures \ref{fig16}(b) and \ref{fig17}(b), except that the dips in Figures \ref{fig16}(b) and \ref{fig17}(b) emerge at around 0.5 pc. If this fraction is taken as an indicator of the importance of self-gravity, this result suggests that gravity becomes increasingly significant from small scales (0.25 pc) to large scales (10 pc). The variation trend does not differ significantly between regions inside and outside the \ion{H}{2} region.

The variation of the virial parameter with mass and radius of substructures has also been investigated for the Maddalena GMC by \cite{Shen2024}. In that companion paper, we found that self-gravity plays a minor role in the quiescent regions of Maddalena at scales below 5 pc, while it becomes important on scales from 0.8 pc to 4 pc in the IRAS 06453 star-forming region. In contrast, our results for the RMC show a different behavior, with the fraction of self-gravitating structures increasing steadily with radius and mass from nearly zero to close to unity with increasing radius and mass. If we adopt a threshold of 50\% for the fraction to indicate gravitational dominance, then self-gravity becomes important on scales larger than 1 pc in the RMC. However, at smaller scales (below 1 pc), the importance of gravity is relatively uncertain. This is because gravitational dominance is not solely determined by size, but also depends on the mass and surface density of the structure. A small structure can still be gravitationally bound if it has a sufficiently high surface density or mass. 

\section{Summary and Conclusion}\label{sec5}

In this work, we used the MWISP $\rm ^{13}CO\  J=1-0$ data to analyze the internal structures of the RMC. To avoid artificial substructures resulting from the binary decomposition of the original Dendrogram algorithm, we developed a non-binary Dendrogram algorithm that allows multiple substructures to sprout from one branch. We decomposed the $^{13}$CO emission from the RMC into 588 substructures in total, including 458 leaves and 130 branches. Physical parameters such as mass, radius, velocity dispersion, and virial parameter of these substructures are calculated. Scaling relationships, such as $\sigma_v-R$, $M-R$, $\sigma_v^2/R-\Sigma$, are investigated using the identified substructures from a hierarchical perspective. We also discussed the physical properties of the gravitationally bound structures and investigated the importance of self-gravity on various spatial scales in the RMC. The main findings of this work are summarized as follows.

\begin{enumerate}
	\item  The substructures identified from the non-binary Dendrogram algorithm show exponential distributions of $T_{\text{peak}}$ and $T_{\text{diff}}$. The fitted exponents of the $T_{\text{peak}}$ and $T_{\text{diff}}$ distributions indicate characteristic peak brightness temperatures and brightness temperature differences that are above 5 $\sigma_{\rm RMS}$, highlighting the intrinsic hierarchy of the $\rm ^{13}CO$ emission in the RMC. Substructures inside the Rosette Nebula tend to exhibit larger $T_{\text{peak}}$ and $T_{\text{diff}}$. 

	\item  The obtained size and mass of the substructures follow power-law distributions with indices consistent with those obtained using traditional structure identification algorithms. Moderate differences are observed between substructures inside and outside the \ion{H}{2} region, with the distributions being generally flatter for structures inside the \ion{H}{2} region.

	\item The velocity dispersions of the structures scale with their radii, exhibiting large scatters and yielding a power-law correlation with an exponent around 0.6 and a relatively low correlation coefficient. However, a more robust correlation is observed between $\log \sigma_v$ and $\log (\Sigma R)$, following a relation of $\sigma_v \propto (\Sigma R)^{0.29\pm0.01}$, with a higher Pearson coefficient of 0.73.
 
	\item Self-gravitating structures are present across a wide range of scales in the RMC, from $\sim$ 0.2 pc to 10 pc. Substructures with higher mass or surface density are more likely to be gravitationally bound. Our results are consistent with those of \cite{Goodman2009}, but extend to a much broader range of scales over which self-gravitating structures exist.

\end{enumerate}

\begin{acknowledgments}
    We thank the anonymous referee for the referee for the constructive comments that help to improve this manuscript. This research made use of the data from the Milky Way Imaging Scroll Painting (MWISP) project, which is a multiline survey in $^{12}$CO/$^{13}$CO/C$^{18}$O along the northern Galactic plane with the PMO-13.7 m telescope. We are grateful to all the members of the MWISP working group, particularly the staff members at the PMO-13.7 m telescope, for their long-term support. MWISP was sponsored by National Key R$\&$D Program of China with grant 2023YFA1608000, 2023YFA1608002, 2017YFA0402701 and by CAS Key Research Program of Frontier Sciences with grant QYZDJ-SSW-SLH047. Y.M. acknowledges the support of NSFC grant 12303033. H.W., Y.M., and M.Z. acknowledge the support of NSFC grant 12473026. X.C. acknowledges the support of NSFC grant 12041305 and the support from the Tianchi Talent Program of Xinjiang Uygur Autonomous Region. This research made use of astrodendro, a Python package to compute dendrograms of astronomical data (http://www.dendrograms.org/). 
	
\end{acknowledgments}

%

\vspace{5mm}




\clearpage
\appendix


\section{The Non-binary Dendrogram Procedure}
\subsection{Modification to the Original Algorithm} \label{appA1}

The astrodendro package, based on the algorithm of \citet{Rosolowsky2008}, constructs a strictly binary hierarchical tree, in which each branch connects just two substructures. This binary constraint does not reflect the observed fragmentation of molecular clouds, where a single structure may split into multiple components. In the original algorithm, any two independent structures (leaves or branches) connected by a single voxel are merged into a new branch, and thresholds such as $min\_delta$ and $min\_npix$ are applied only to leaves, not to branches, allowing the branches to exist even if their brightness temperature differences are insignificant. This often leads to the appearance of ``phantom branches'' \citep{Storm2014}, which are embedded within the dendrogram but lack sufficient brightness temperature difference to represent true hierarchical levels. For example, their $T_{\text{diff}}$ could be much smaller than the noise level.

To address this issue, the following changes have been made to the original astrodendro: when two structures are about to merge, the branch structures are no longer considered to be independent if they do not satisfy the minimum voxel count ($min\_npix$) and the minimum brightness temperature difference ($branch\_delta$) requirements. When a dependent branch structure meets an independent leaf structure, the leaf structure will be absorbed into the branch structure as a new sprout in the dendrogram tree. Only when a branch structure meets the conditions $min\_npix$ and $branch\_delta$, can it merge with other independent structures to form a lower-level hierarchical branch.

\subsection{Comparison with the Original Algorithm} \label{app A2}
In this section, we compare the statistical distributions and correlations of properties of hierarchical structures derived with the original astrodendro algorithm and the modified non-binary Dendrogram. The parameters for the original algorithm are set to be the same as those we used for the non-binary Dendrogram, which are $min\_value = 3\ \sigma_{\rm RMS}$, $min\_delta = 2\ \sigma_{\rm RMS}$, and $min\_npix = 27$. Identical approaches are used to calculate the physical properties of these substructures.

Panels (a) and (b) of Figure \ref{fig13} present the {$T_{\text{peak}}$ and $T_{\text{diff}}$ histograms of the hierarchical structures identified using the original algorithm. In comparison with Figure \ref{fig4}, the fitted expotential exponents exhibit only negligible differences from those obtained with the non-binary Dendrogram. The main difference lies in the number of structures within the bins at the smallest $T_{\text{peak}}$ and $T_{\text{diff}}$ scales, indicating that the non-binary implementation effectively suppresses the formation of small spurious branches.

Figure \ref{fig13}(c) and Figure \ref{fig13}(d) show the radius and mass histograms of the structures identified by the original Dendrogram algorithm. The numbers of structures in the largest bins are significantly higher than those in the non-binary case shown in Figure \ref{fig5}. This is possibly due to the increasing number of branches caused by the generation of parental phantom branches. The power-law distributions are also shallower compared to those derived with the non-binary Dendrogram, especially for the radius distribution, where the exponents increase from $-2.43 \sim -2.28$ (in the non-binary case) to $-1.86 \sim -1.84$. This difference is likely a consequence of the presence of phantom branches, which tend to be located in the lower part of the tree and therefore form relatively large and massive spurious structures.

\begin{figure*}[!htb]
	\gridline{\fig{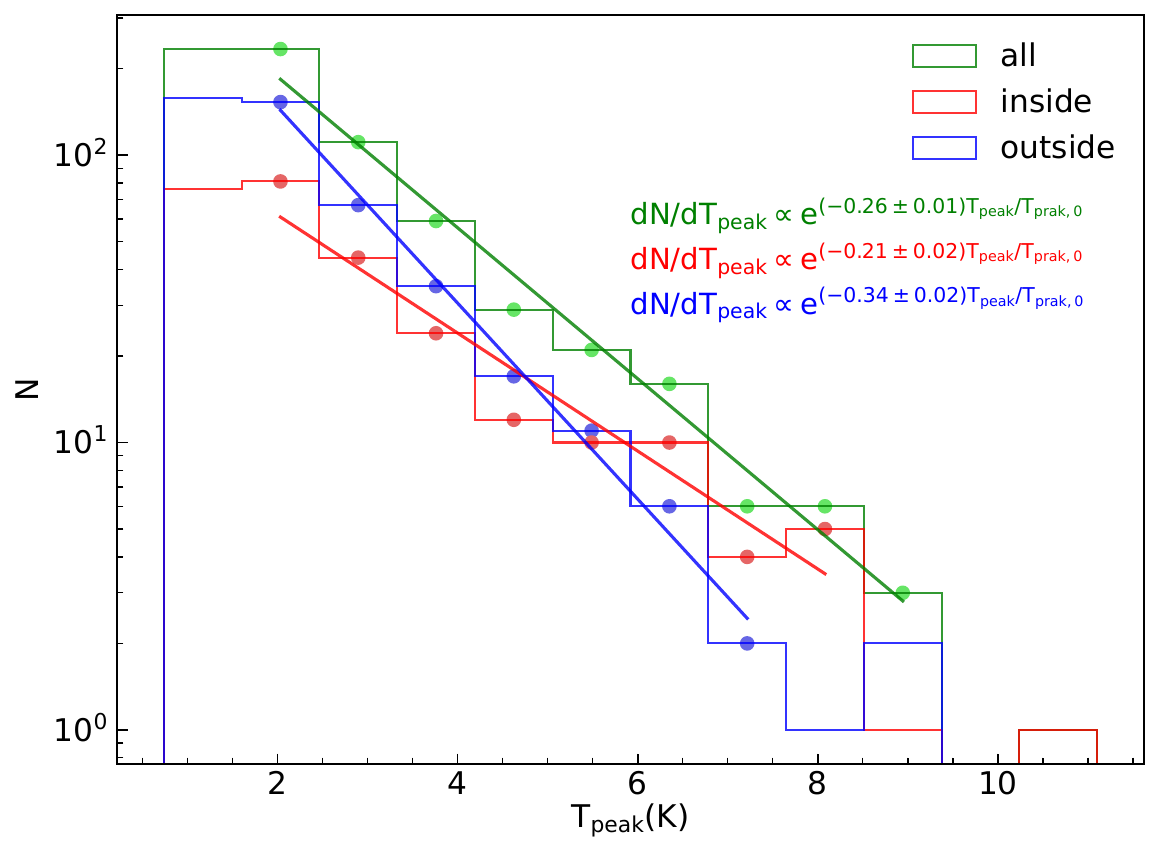}{0.5\columnwidth}{(a)}
	\fig{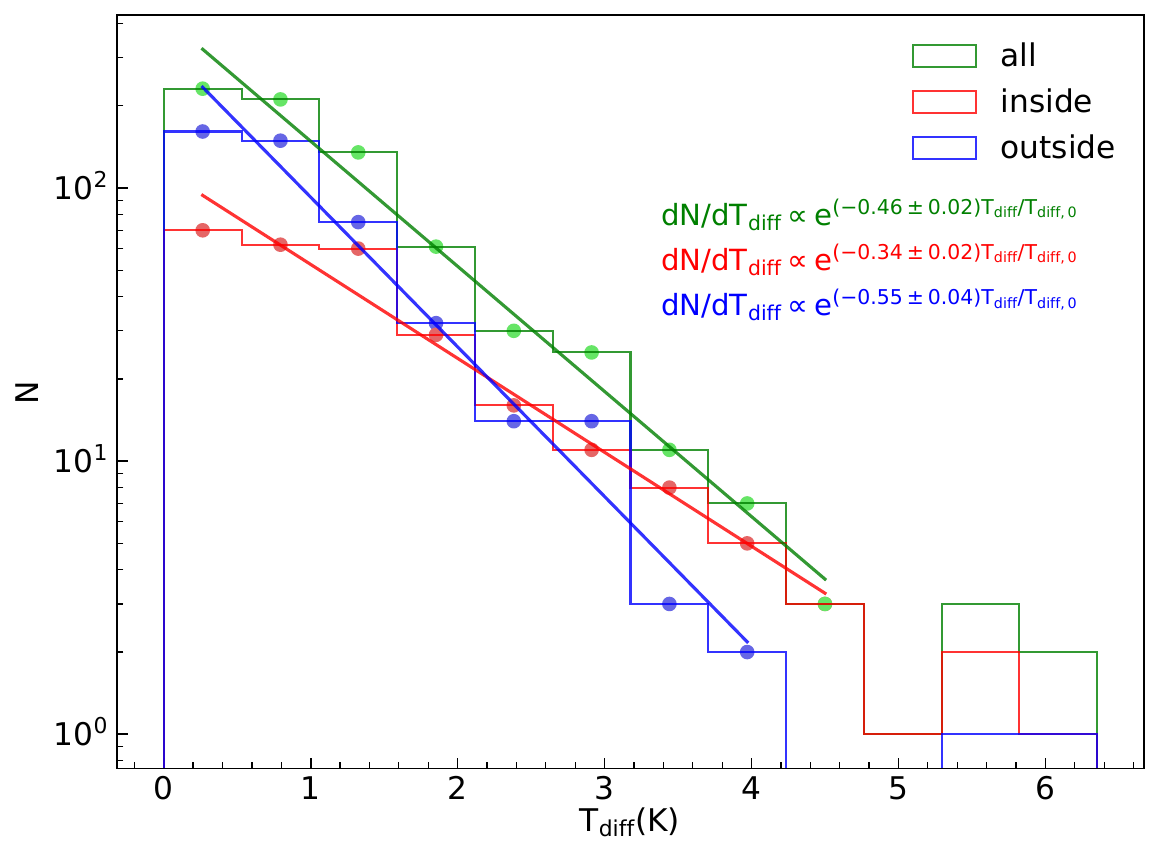}{0.5\columnwidth}{(b)}}
	\gridline{\fig{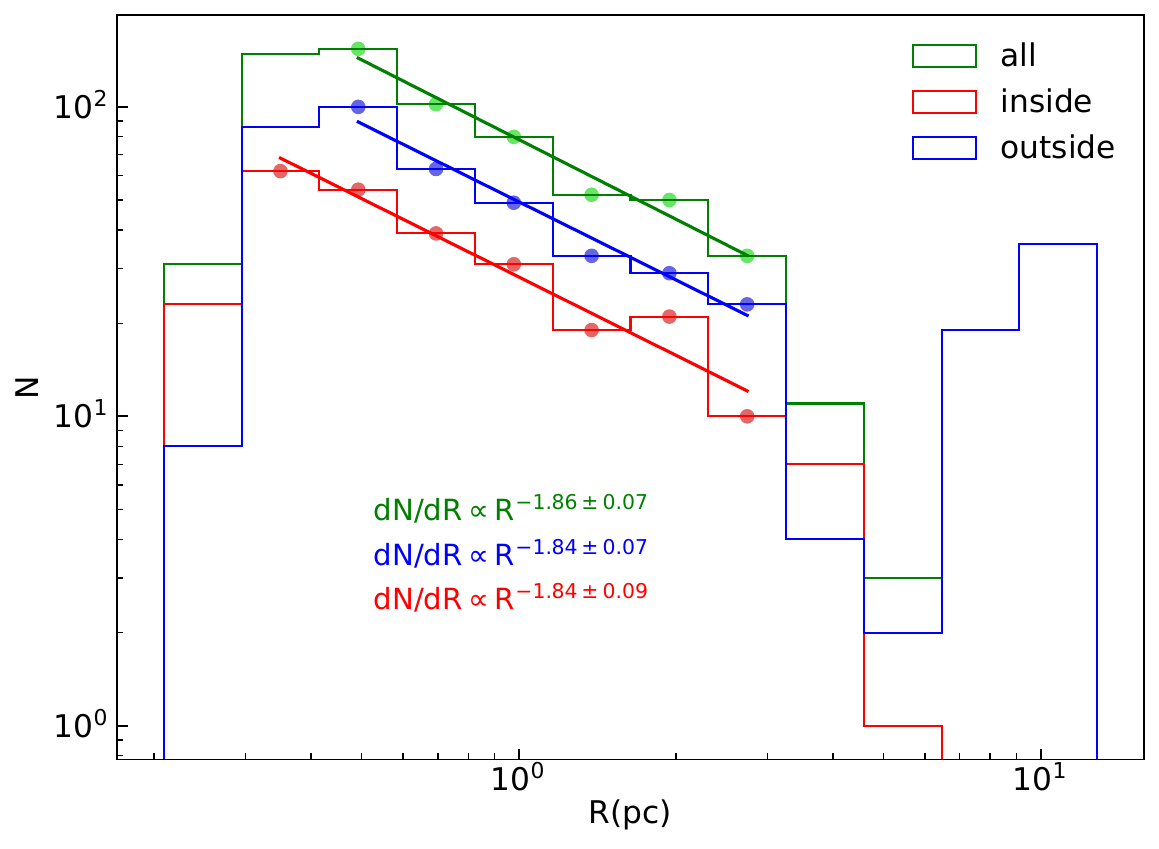}{0.5\columnwidth}{(c)}
	\fig{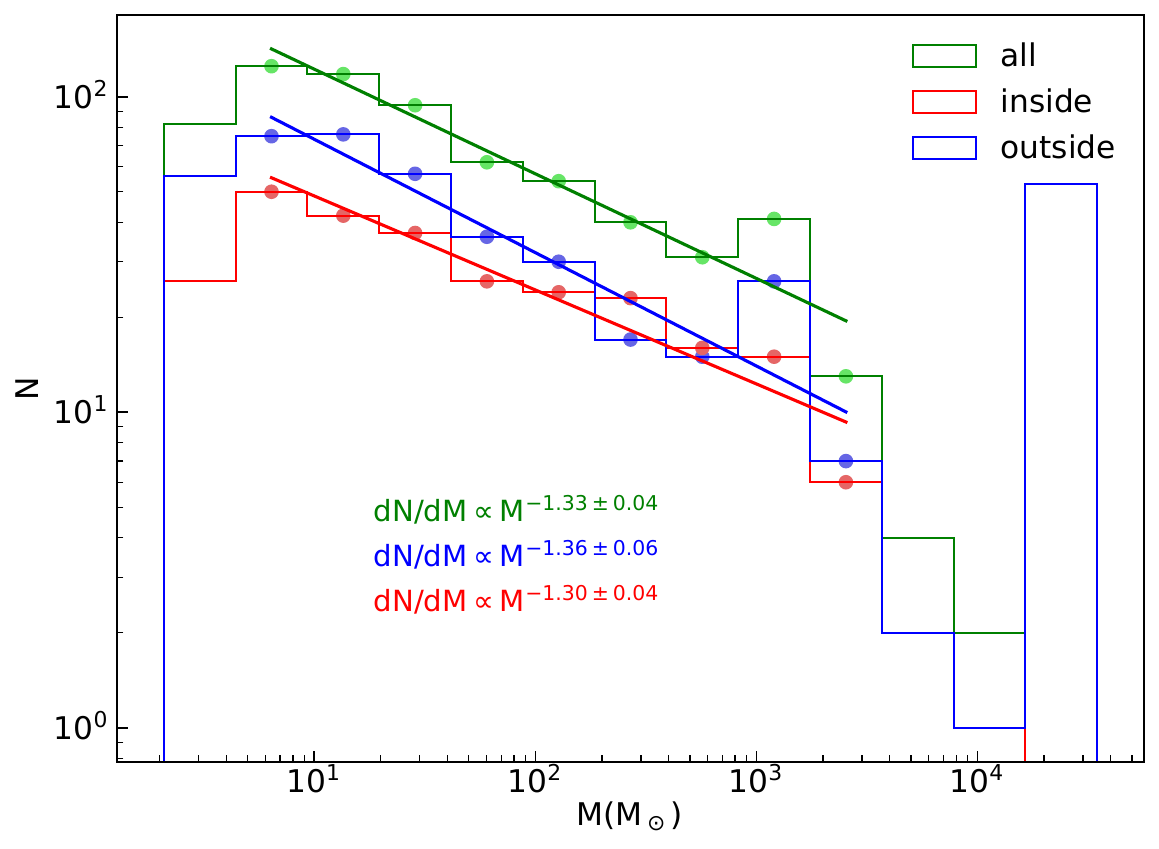}{0.5\columnwidth}{(d)}}
\caption{Histograms of (a) $T_{\text{peak}}$, (b) $T_{\text{diff}}$, (c) $R$, and (d) $M$ of hierarchical structures identified by the original Dendrogram algorithm. The green, red, and blue colors correspond to all structures, structures inside the \ion{H}{2} region, and those outside the \ion{H}{2} region, respectively.} \label{fig13}
\end{figure*}

Figure \ref{fig14} shows the $\sigma_v - R$ relation and the $M-R$ relation of hierarchical structures identified by the original Dendrogram. The power-law indices of both relationships are steeper than those obtained using the non-binary Dendrogram method shown in Figure \ref{fig6}. The power-law slope of $\sigma_v - R$ relation increases from $0.45\pm0.02$ in the non-binary Dendrogram case to $0.54\pm0.02$, while the variation of the power-law index for the $M-R$ relation is relatively small. These variations are also likely caused by the phantom branches, which lead to an increase in the number of points at the upper-right corner in both plots, thereby raising their weights in the fitting. The smaller variation in the index of the $M-R$ relation is possibly due to the fact that the mass distribution has a larger dynamic range compared to the velocity dispersion distribution, therefore, the presence of phantom branches has less effect on the fitting of the relationship.

In conclusion, we suggest that the modifications made to the astrodendro package effectively eliminate the influence of phantom branches, while having little impact on the bright structures. These improvements are crucial for accurately characterizing the statistical properties and scaling relations of hierarchical structures within molecular clouds.

\begin{figure*}[!htb]
	\gridline{\fig{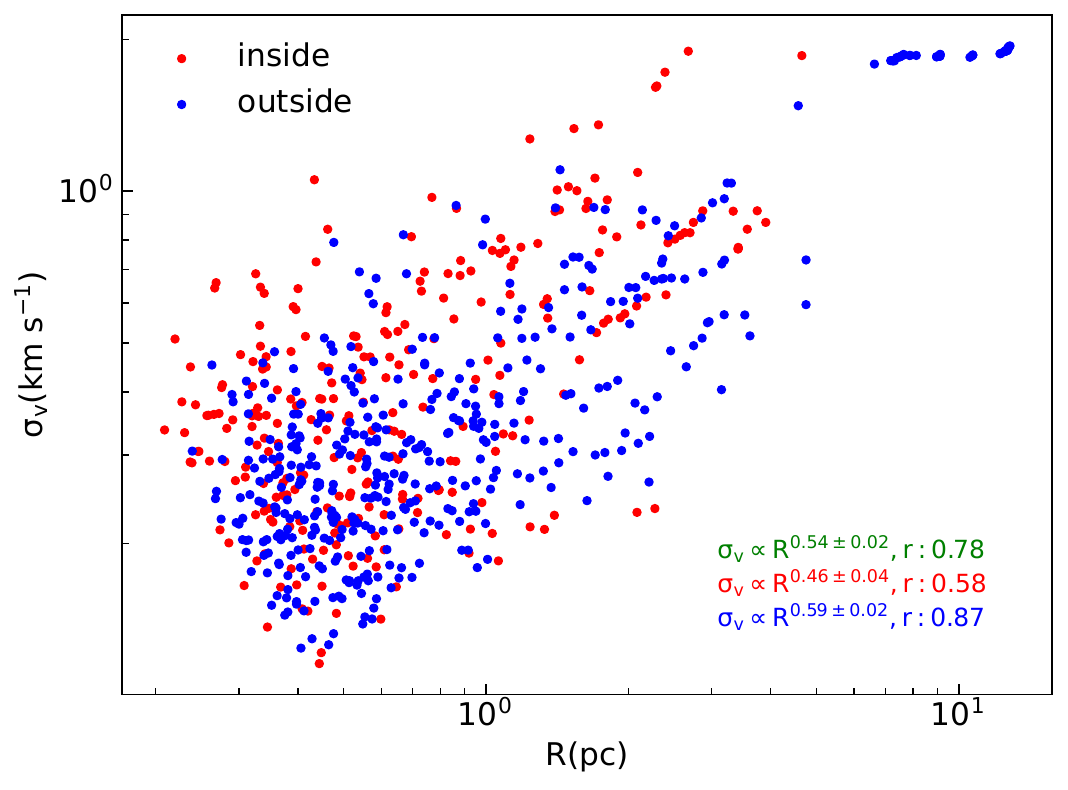}{0.5\columnwidth}{(a)}
	\fig{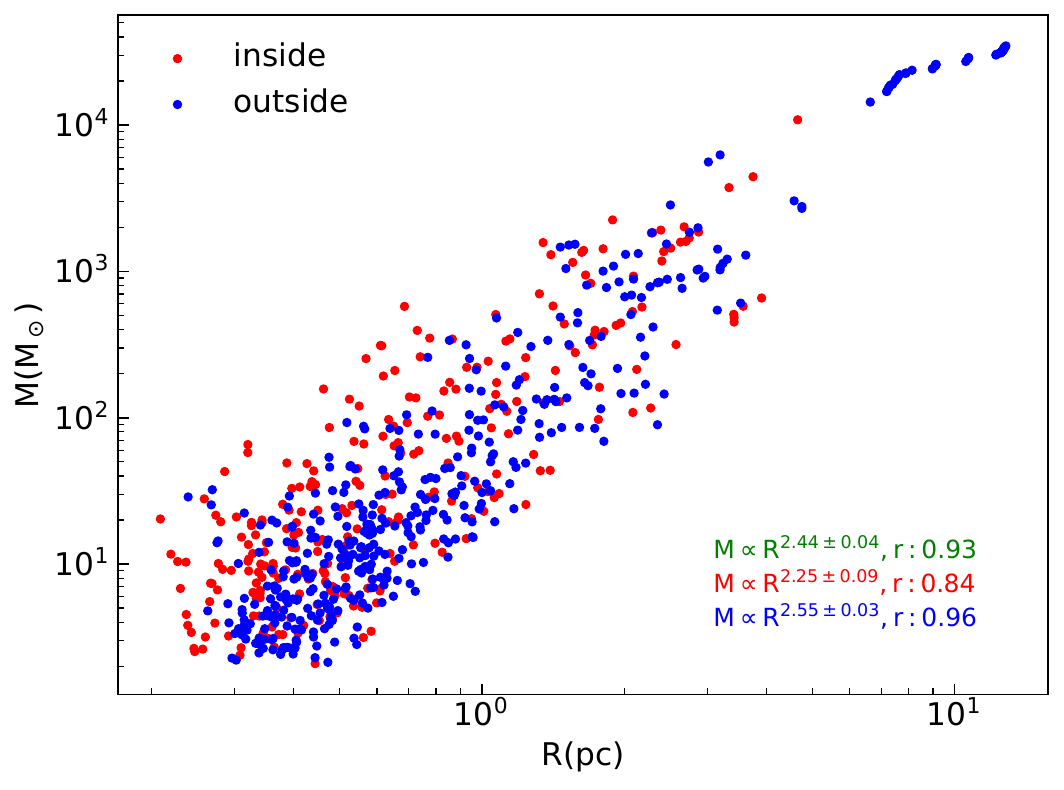}{0.5\columnwidth}{(b)}}
\caption{Relationships between (a) radius and velocity dispersion, and (b) radius and mass for hierarchical structures identified with the original Dendrogram algorithm. Red color and blue color indicate structures inside and outside the \ion{H}{2} region, respectively.} \label{fig14}
\end{figure*}

\section{Influence of Parameter Selection}
\subsection{Influence of Parameter Selection on Scaling Relations} \label{secB1}

When using structure-identification algorithms such as Dendrogram, an inevitable question is how the choice of parameters influences the results. In this work, we adopted a minimum brightness temperature threshold of $3\ \sigma_{\rm RMS}$, a minimum brightness temperature difference constraint for leaves of $2\ \sigma_{\rm RMS}$, a minimum size of 27 voxels for structures to be considered independent, and an additional brightness temperature difference constraint for branches of $1\ \sigma_{\rm RMS}$—which is not present in the original Dendrogram algorithm. An assessment of the robustness of the structure identification with different constraint parameters is necessary.

\begin{figure*}[!htb]
	\gridline{\fig{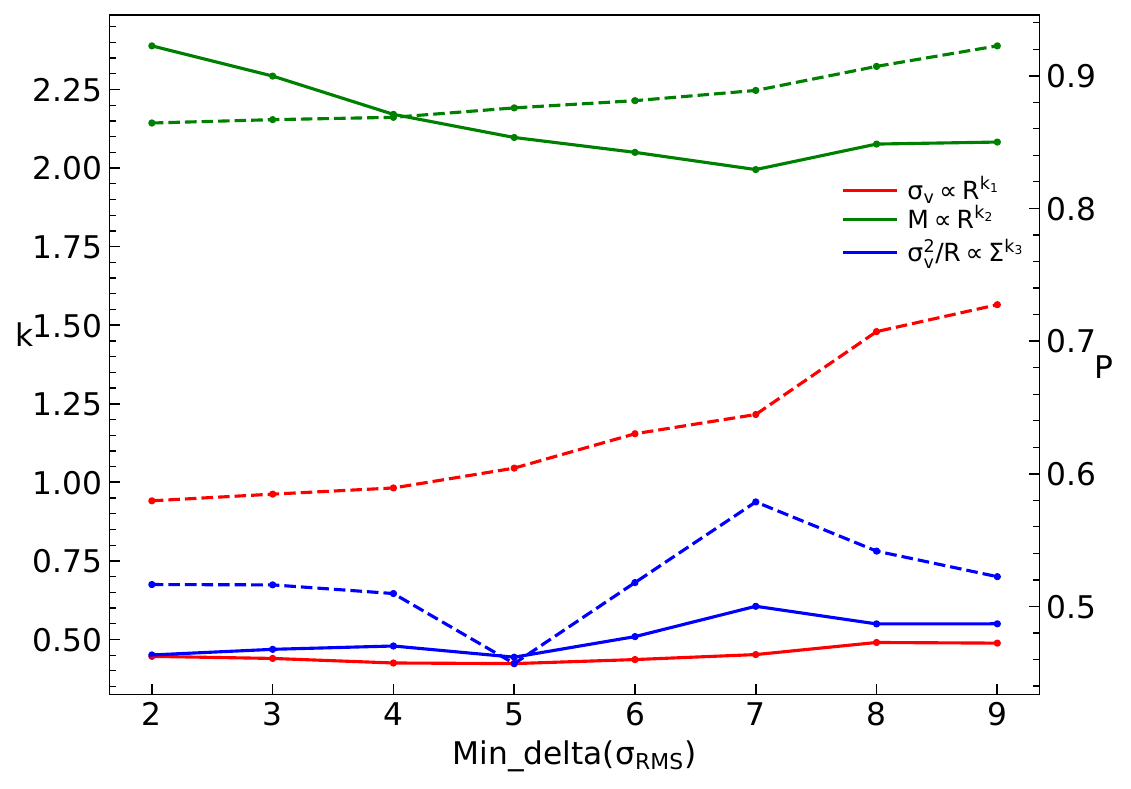}{0.5\columnwidth}{(a)}
	\fig{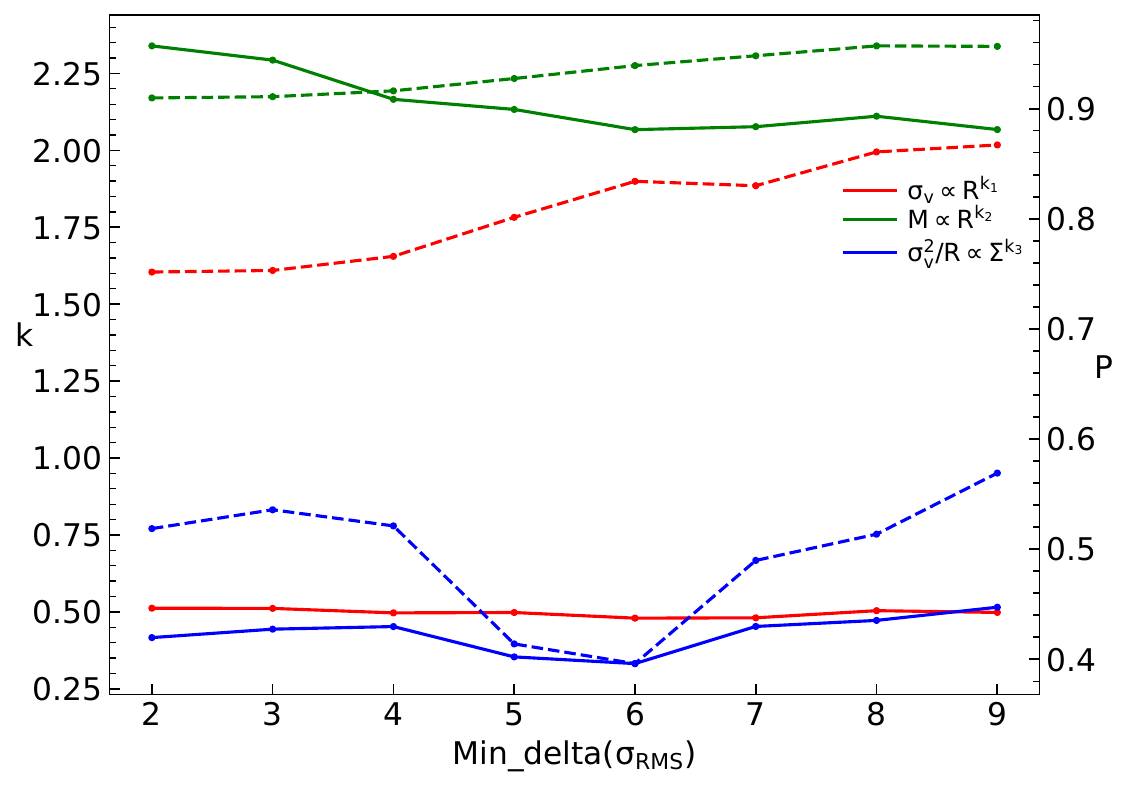}{0.5\columnwidth}{(b)}}
	\gridline{\fig{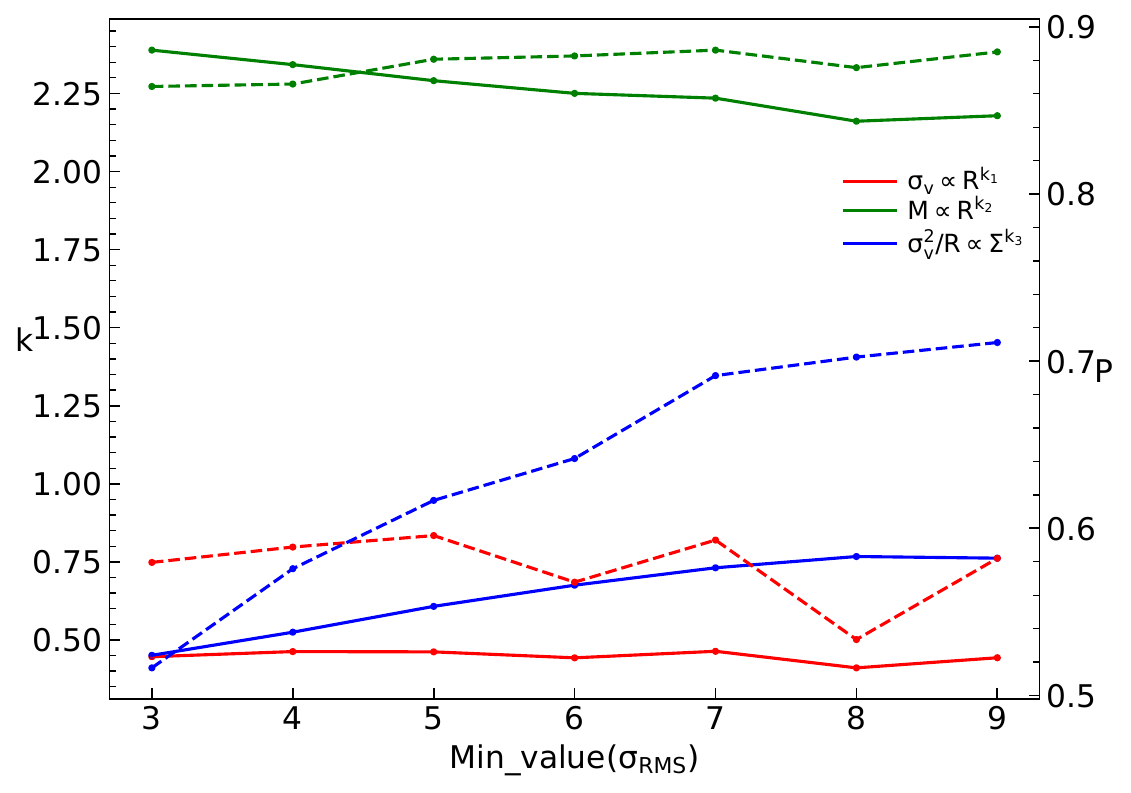}{0.5\columnwidth}{(c)}
	\fig{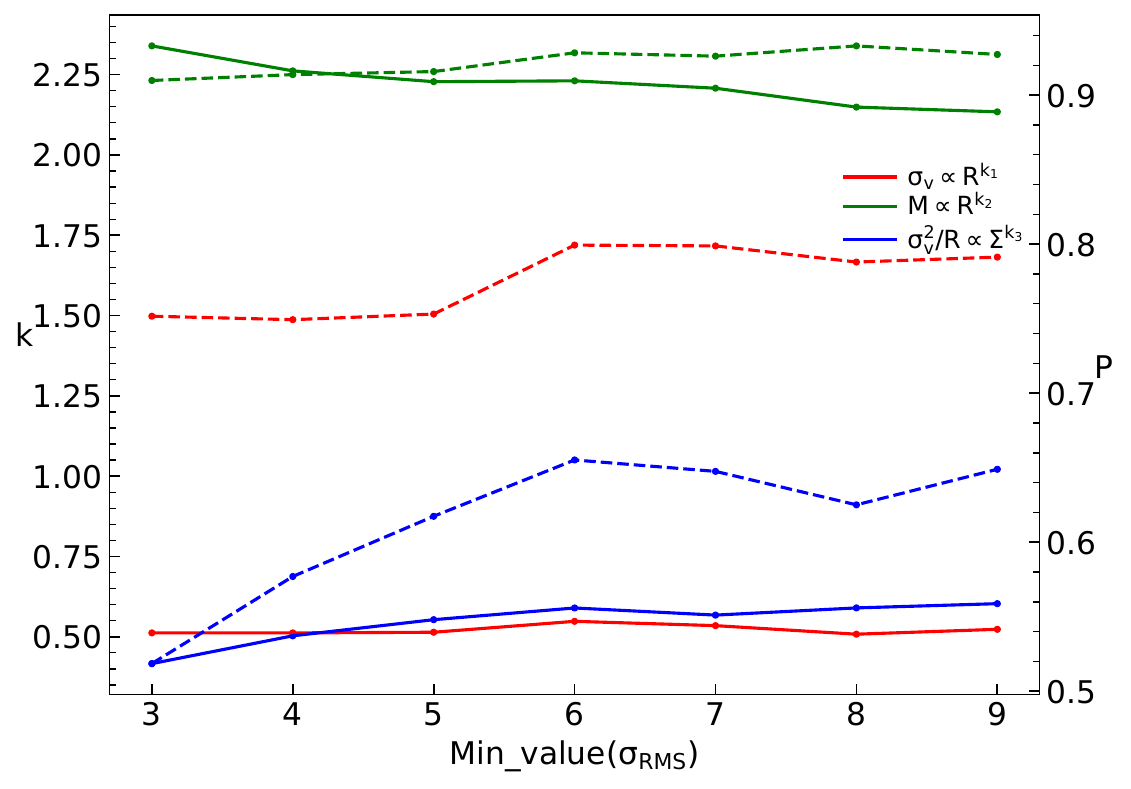}{0.5\columnwidth}{(d)}}
\caption{Variation of the fitted power-law indices in the $\sigma_v-R$, $M-R$, and $\sigma_v^2/R-\Sigma$ relations against $min\_delta$ for (a) substructures in the entire RMC and (b) the biggest tree shown in Figure \ref{fig3}. Panels (c) and (d) show the same variations but against $min\_value$ instead. Different relations are distinguished by colors, with the solid lines representing the power-law index of each relation (corresponding to the left vertical axis), and the dashed lines indicating Pearson's correlation coefficients (corresponding to the right vertical axis).} \label{fig15}
\end{figure*}

We run the modified non-branch Dendrogram on RMC with increasing $min\_delta$ by a step of $\sigma_{\rm RMS}$ to test the behavior of the scaling relations discussed in Section \ref{sec3}. Simultaneously, we examine the scaling relations for substructures within the largest tree only, because substructures within a single tree are expected to be inherently hierarchical. 

Figure \ref{fig15} shows the variation of the fitted power-law indices, $k_i$, in the relations of $\sigma_v \propto R^{k_1}$, $M \propto R^{k_2}$, and $\sigma_v^2/R \propto \Sigma^{k_3}$ against the input parameters $min\_delta$ and $min\_value$. Different relationships are represented by different colors, while solid lines indicate the values of power-law exponents (corresponding to the left y-axis), and dashed lines represent the corresponding Pearson correlation coefficients (corresponding to the right y-axis). The fitted indices $k_1$ for both substructures in the entire RMC and in the biggest tree are generally stable around 0.5 against different $min\_delta$. A systematic larger correlation coefficient between $\log \sigma_v$ and $\log R$ is observed when considering substructures within the largest tree only, as demonstrated by the red dashed lines in panels (a) and (b) of Figure \ref{fig15}. This likely reflects stronger physical connection among substructures in the largest tree.
By contrast, the other two scaling relations show no significant difference between all structures and those within the largest tree. The power-law index $k_2$ in the $M-R$ relation gradually decreases with $min\_delta$ and asymptotically reaches the value of 2. The correlation coefficient of the $M-R$ relation is the highest among the three scaling relations for all values of $min\_delta$. For the relation between $\sigma_v^2/R$ and  $\Sigma$, the power-law index $k_3$ remains near 0.5 across all $min\_delta$ values. Its Pearson correlation coefficient reaches a minimum value (0.4-0.5) when $min\_delta$ is set to $5\ \sigma_{\rm RMS}$ (panel (a)) or $6\sigma_{\rm RMS}$ (panel (b)). This contrasts with the other two correlation relationships, where the coefficients increases monotonically with $min\_delta$ throughout the tested parameter range. We can see that the $\sigma_v-R$, $M-R$, and the $\sigma_v^2/R-\Sigma$ relations are stable against the selection of the parameter $min\_delta$ used in our non-binary Dendrogram algorithm.

The same examinations are also implemented for different values of the parameter $min\_value$, as shown in Figures \ref{fig15}(c) and (d). The index $k_1$ remains stable at around 0.5 for increasing $min\_value$, with significantly improved correlation coefficients for structures within the biggest tree. The index $k_2$ and the correlation coefficient of the $M-R$ relation show little change with $min\_value$ in panels (c) and (d). For the relation between $\sigma_v^2/R$ and $\Sigma$, the $k_3$ systematically increases with increasing $min\_value$. The correlation coefficient in this relation also increased in general with $min\_value$. The three relations examined in our test show robustness against the selection of the parameter $min\_value$.

\subsection{Influence on Virial Parameter} \label{secA2}

\begin{figure*}[!htb]
	\centering
	\begin{minipage}[ht]{0.8\columnwidth}
		\centering
		\includegraphics[trim = 0cm 0cm 0cm 0cm, width=\columnwidth, clip]{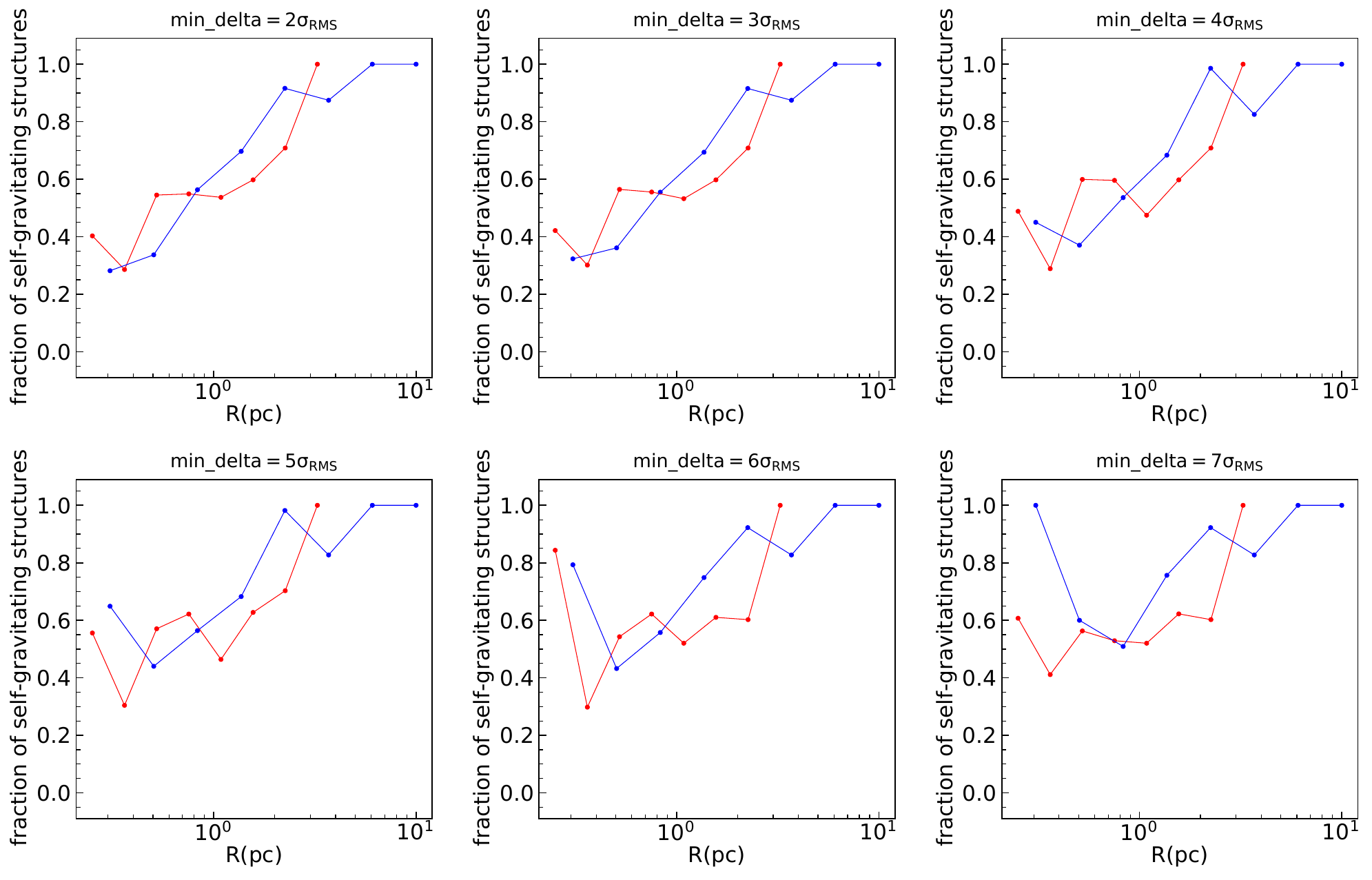}
		\\{(a)}
	\end{minipage}
	\begin{minipage}[ht]{0.8\columnwidth}
		\centering
		\includegraphics[trim = 0cm 0cm 0cm 0cm, width=\columnwidth, clip]{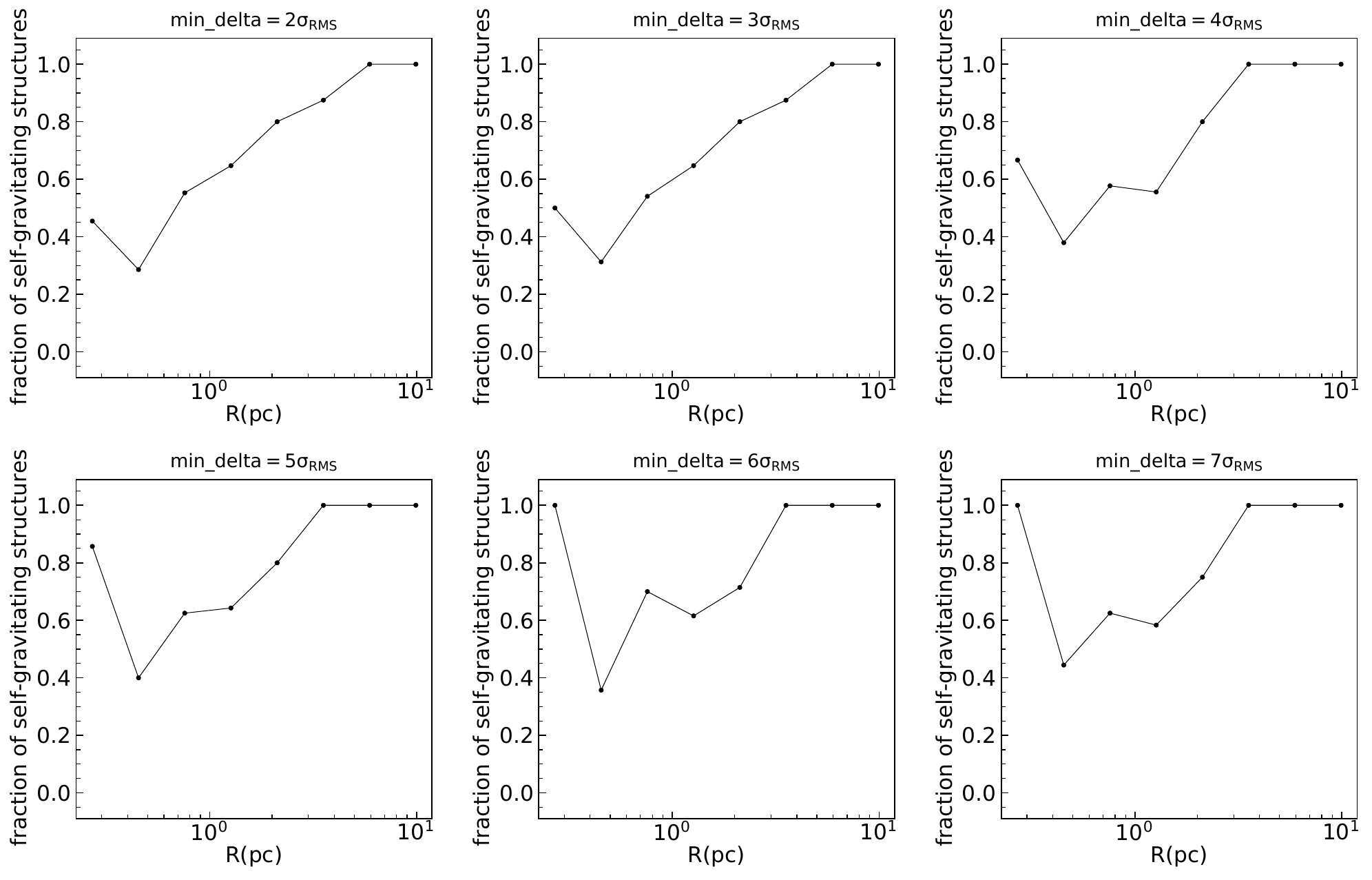}
		\\{(b)}
	\end{minipage}	

\caption{Variation of the fraction of self-gravitating structures against radius derived from test runs with increasing $min\_delta$ for substructures in (a) the entire RMC and (b) the biggest tree shown in Figure \ref{fig3}. Substructures inside and outside the \ion{H}{2} region are highlighted with red color and blue color, respectively. }\label{fig16}
\end{figure*}

\begin{figure*}[!htb]
	\centering
	\begin{minipage}[h]{0.8\columnwidth}
		\centering
		\includegraphics[trim = 0cm 0cm 0cm 0cm, width=\columnwidth, clip]{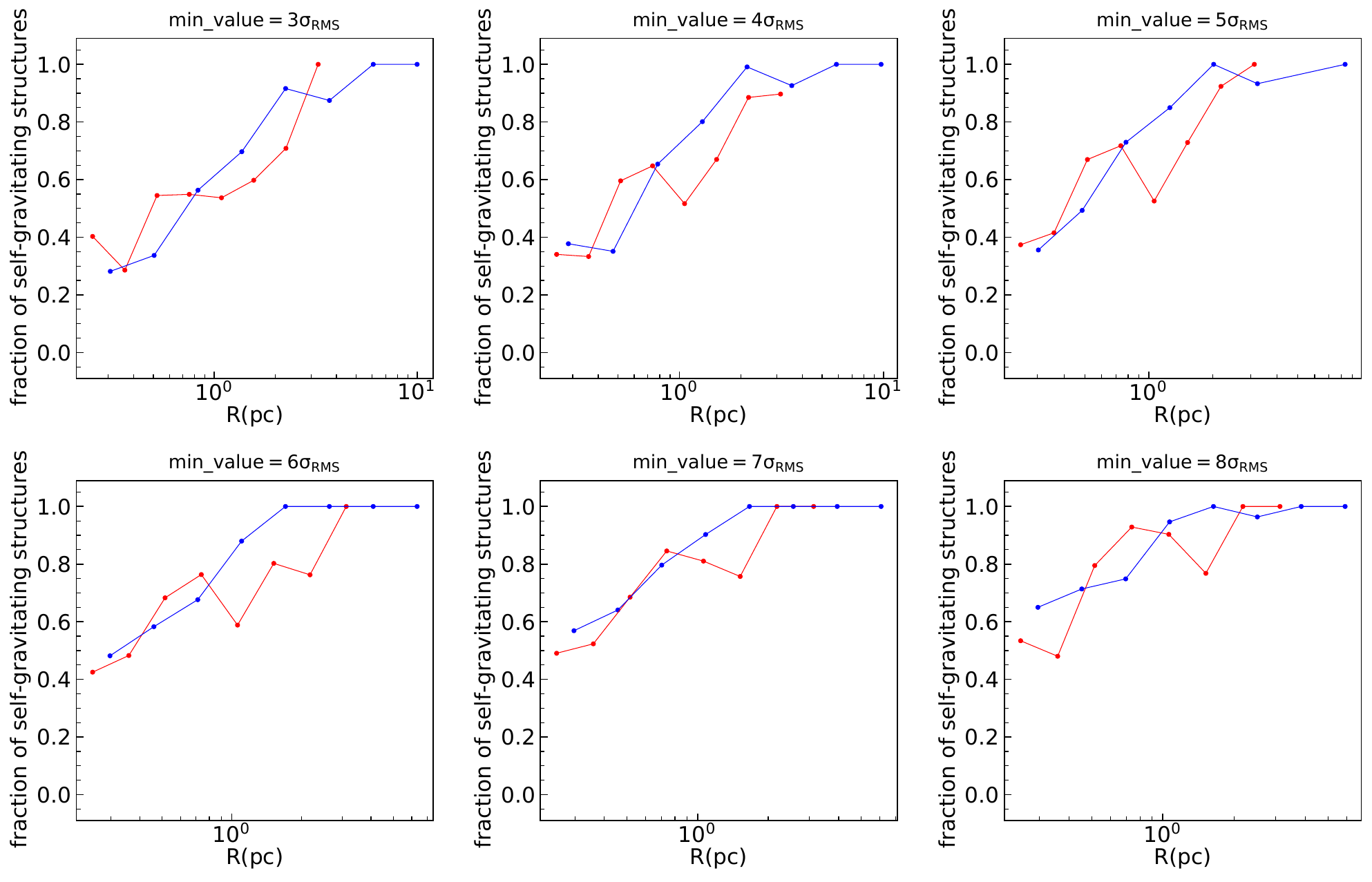}
		\\{(a)}
	\end{minipage}
	\begin{minipage}[h]{0.8\columnwidth}
		\centering
		\includegraphics[trim = 0cm 0cm 0cm 0cm, width=\columnwidth, clip]{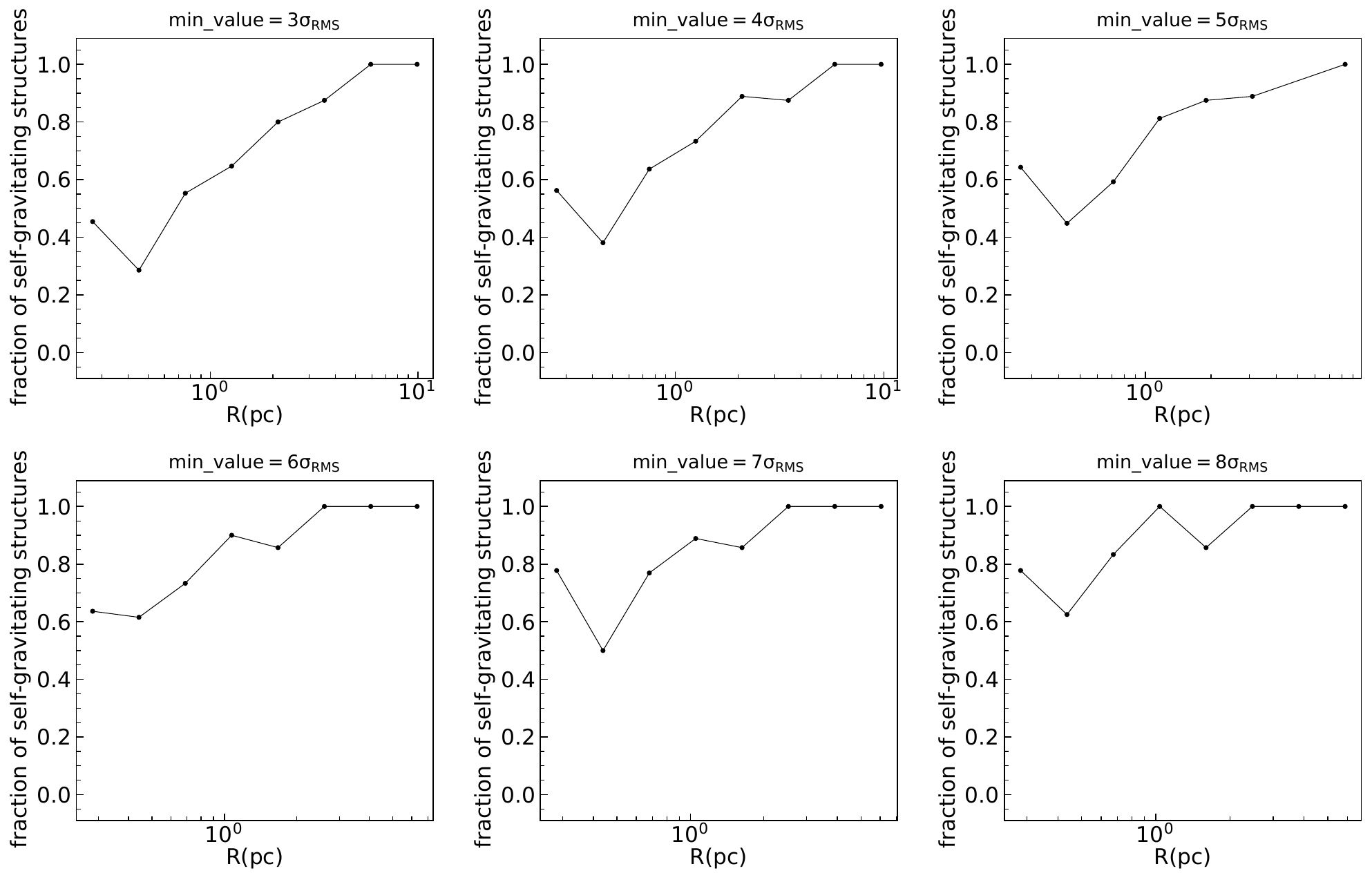}
		\\{(b)}
	\end{minipage}	

\caption{Same as Figure \ref{fig16}, but for test runs with different $min\_value$. }\label{fig17}
\end{figure*}

We tested the variation of the fraction of self-gravitating structures against radius and mass with different $min\_delta$ and $min\_value$.
Figure \ref{fig16} shows the variation of the fraction of self-gravitating structures with the parameter $min\_delta$. The parameter $min\_value$ is fixed to $3\sigma_{\rm RMS}$ in these test runs. In panel (a), different colors are used to annotate structures inside and outside the \ion{H}{2} region. Panel (b) shows the relation for substructures within the biggest tree. 
As $min\_delta$ increases, the fraction in the smallest radius bin shows a gradual rise. In Figures \ref{fig16}(a) and (b), the curves corresponding to $min\_delta = 5\ \sigma_{\rm RMS}$, $6\ \sigma_{\rm RMS}$, and $7\ \sigma_{\rm RMS}$ exhibit a dip. Overall, the variation of the fraction o f self-gravitating structures with radius is stable against the parameter $min\_delta$.

Figure \ref{fig17} shows the variation of the fraction of self-gravitating structures with respect to the parameter $min\_value$, i.e., the minimum brightness temperature threshold. Similar to the test on $min\_delta$, we fixed $min\_delta$ at $2\sigma_{\rm RMS}$ in the test runs in Figure \ref{fig17}. As tested with the $min\_delta$ parameter, the variation of the fraction of self-gravitating structures with structure radius show strong robustness with the selection of the parameter $min\_delta$. Our test runs confirm that the monotonically increasing fraction of self-gravitating substructures obtained in Figure \ref{fig10} is robust to a certain extent against the input parameters of the algorithm.


\clearpage
\bibliography{reference}{}
\bibliographystyle{aasjournal}



\end{document}